\documentclass[sensors,article,accept,moreauthors,pdftex,10pt,a4paper]{Definitions/mdpi}

\usepackage[utf8]{inputenc} 
\usepackage[T1]{fontenc}    
\usepackage{amsfonts}       
\usepackage{nicefrac}       
\usepackage{bm}
\usepackage{subcaption}

\firstpage{1} 
\pubvolume{xx}
\issuenum{1}
\articlenumber{1}
\pubyear{2019}
\copyrightyear{2019}
\externaleditor{Academic Editor: name}
\history{Received: date; Accepted: date; Published: date}

\Title{Centre of pressure estimation during walking using only inertial-measurement units and end-to-end statistical modelling}

\newcommand{\etal}{\textit{et al.}}

\Author{Janez Podobnik$^{1,\dagger}$*\orcidB{}, David Kraljić $^{1,\dagger}$\orcidA{}, Matjaž Zadravec$^{2}$ and Marko Munih$^{1}$}

\AuthorNames{David Kraljić, Janez Podobnik, Matjaž Zadravec and Marko Munih}

\address{%
$^{1}$ \quad University of Ljubljana, Faculty of Electrical Engineering;  \{janez.podobnik, david.kraljic, marko.munih\}@fe.uni-lj.si\\ 
$^{2}$ \quad University Rehabilitation Institute Republic of Slovenia; matjaz.zadravec@ir-rs.si}

\corres{Correspondence: janez.podobnik@fe.uni-lj.si}

\firstnote{These authors contributed equally to this work.} 

\abstract{Estimation of the centre of pressure (COP) is an important part of the gait analysis, for example, when evaluating the functional capacity of individuals affected by motor impairment. Inertial measurement units (IMUs) and force sensors are commonly used to measure gait characteristic of healthy and impaired subjects. We present a methodology for estimating the COP solely from raw gyroscope, accelerometer, and magnetometer data from IMUs using statistical modelling. We demonstrate the viability of the method using an example of two models: a linear model and a non-linear {Long-Short-Term Memory} (LSTM) neural network model. Models were trained on the COP ground truth data measured using an instrumented treadmill and achieved the average intra-subject {root mean square (RMS) error between estimated and ground truth COP} of 12.3\,mm and the average inter-subject RMS error of 23.7\,mm which is comparable or better than similar studies so far. 
We show that the calibration procedure in the instrumented treadmill can be as short as a couple of minutes without the decrease in our model performance. We also show that the magnetic component of the recorded IMU signal, which is most sensitive to environmental changes, can be safely dropped without a significant decrease in model performance. Finally, we show that the number of IMUs can be reduced to five without deterioration in the model performance.}

\begin{document}



\section{Introduction}
\label{sec:intro}
Walking is a major part of daily living activities and is, therefore, an important aspect of autonomy. Gait analysis of both healthy and impaired individuals gives valuable information about their functional capacity. For stroke patients, for example, the analysis contributes to the selection of proper rehabilitation training \cite{doi:10.1161/STROKEAHA.109.558940, Caldas2017}. COP estimation, in addition to various other parameters, is an important step in biomechanical analysis of gait during level walking \cite{Roberts2017, Ancillao2018, 4957058}.

Most commonly, COP is measured using force plates or force sensors embedded in the floor or a walking platform. Gait laboratories are typically equipped with one, two, or three force plates which makes tracking successive steps during walking over multiple steps impossible \cite{WINTER1995193, s17010075, Shahabpoor2017}. To solve this issue researchers either use sensors which are embedded into instrumented shoes or insoles \cite{6471220}, or by measuring gait on an instrumented treadmill with force sensors embedded into the treadmill \cite{6145657}.

Another approach in studying the COP is to estimate it by measuring the kinematics during walking. Winter \cite{WINTER1995193} has proposed a simple model to calculate the COP from the centre of mass (COM). COM is calculated using a kinematic method. In the kinematic method, accurate anthropometric model and full kinematic description of each segment are required. Audu \etal \cite{AUDU20071115} have developed a model for estimating ground reaction force (GRF) and COP during standing from kinematic data of the human body segments. Ren \etal \cite{REN20082750} presented a whole-body inverse dynamics model based only on measured kinematics for gait, however, the model was used to predict GRFs and not the COP.  A similar method based on the human dynamic model and kinematic measurements of human body segments were recently presented in  \cite{s17010075,JUNG201662}. The method was primarily developed for the estimation of GRFs and joint moments, however, the authors have also estimated the COP under each foot and compared it to the reference COP measured using the force plates.

Artificial neural networks (ANN) are a flexible tool for statistical modelling of nonlinear processes in general, but also in the analysis of gait \cite{SCHOLLHORN2004876}. An important feature of ANN is that information is stored in the form of weights, which are set during training using examples (training set of inputs and outputs) from which the ANN networks learn so that the information about the relation between inputs and outputs can be used to predict the outputs given new inputs. Funk \etal \cite{Funk2019} developed a method based on convolutional neural networks to estimate the COP from a video of a subject performing movements. Pose was extracted from the video with corresponding foot pressure maps used to train the neural network. Most commonly, ANNs are used as a calibration function which maps raw sensor data from force or pressure sensors into the estimated COP \cite{HSIEH2016217, 8365154, s18124349}. 

The majority of studies use an optical motion capture system, which is considered a gold standard for gait analysis. However, the recent development of more accurate, stable and smaller IMUs allows the estimation of human body kinematics with satisfactory accuracy \cite{Caldas2017, armIMUreview, 7403983}. Compared to optical motion capture systems, which achieve better accuracy,  IMUs are low-cost and portable and hence more practical for use in non-laboratory setting \cite{armIMUreview, diseases7010018}. IMUs are therefore often used for gait analysis. Majority of studies that use IMUs in gait assessment use one or up to 5 units \cite{diseases7010018}. Most commonly, IMUs are used for measuring joint angles, spatio-temporal parameters, for detection of gait events, and for classification of different pathological gaits \cite{Roberts2017, Caldas2017, diseases7010018}.

\paragraph{Motivation}
The main motivation for this study was to develop a methodology to estimate the COP during
walking solely from IMUs. IMU technology was selected as it enables the estimation of the global
COP and thus an easy way of detecting the asymmetry in the COP ``butterfly'' shape
\cite{ROERDINK20082628, Roerdink2014}, which is relevant for analysis of pathological gait (e.g.
stroke patients \cite{zadrnovel}, amputee patients \cite{ Roerdink2014}, patients with Parkinson's
disease \cite{diseases7010018}, children with Angelman syndrome \cite{Grieco2018}). Global COP (sometimes referred also as the gaitogram \cite{Roerdink2014}) is especially important since it allows to measure not just temporal gait parameters {such as stride time, step time, single- and double-support stance durations, but also spatial gait parameters such as step width, step length, single-stance width, single-stance length and stride length \cite{VERKERKE20051881, ROERDINK20082628,Grieco2018}. The spatio-temporal gait characteristics obtained from COP pattern may indicate different information about gait balance, e.g. larger step width obtained from the mediolateral COP displacement indicate larger mediolateral stability, while step time and single stance length obtained from the anteroposterior COP displacement determine the duration of loading body weight to the lower extremity. Thus, by obtaining the COP from an instrumented treadmill or estimating it during level walking, a valuable knowledge is given for studying the balance of gait, and can be used for assessment of persons with
movement disabilities \cite{ROERDINK20082628}.} Surprisingly, little amount of research covers the
estimation of the COP using IMUs \cite{Ancillao2018}. Only a small number of studies have been
done to estimate the COP from IMUs only \cite{Shahabpoor2017}. Studies presented in
\cite{s17010075,JUNG201662} used a dynamic model for the human body to estimate the COP.

\paragraph{Our contribution \& aims}
The main novelty of our contribution is the development and analysis of the method for estimating the global COP during gait solely from IMUs using statistical modelling. The method has been analysed (a) in terms of accuracy and
compared to results from other studies related to estimation of COP, (b) how number and
constellation of IMUs affects the accuracy of the estimation of the COP, (c) how the amount of the
training data affects the performance of the models, and (d) how the models trained on one set of
subjects generalise and transfers to a different subject.

The goal of this study is to demonstrate a method for estimating the COP during level walking from
wearable IMUs using end-to-end statistical modelling and to evaluate how well the IMU technology
performs with our method. We performed measurements on six subjects wearing IMUs while
walking in an instrumented treadmill that recorded the COP. In addition to our measurements, we
also make use of a public dataset \cite{synthetic} of walking kinematics. We restructure the public
dataset such that the data is in the same form as our measurements. This way we expanded the
number of subjects on which we evaluated our methods in order to provide more support for the
results obtained on our dataset.

\paragraph{}
This paper is organised as follows: Section \ref{sec:intro} introduces the problem of estimating the COP using IMUs, related work, and presents the motivation for the presented study in this paper. Section \ref{sec:methods} describes the experimental protocol and proposed method for end-to-end mapping between the raw IMU signal and the COP during walking. Section \ref{sec:results} describes the results of experiments in terms of accuracy of the estimated COP compared to the ground truth COP measured using an instrumented treadmill and discusses the impact of reducing the number of used IMUs, how to reduce time spent acquiring training data, and model transferability between different subjects. Section \ref{sec:conclusion} presents our conclusions and future work.

\section{Material and Methods}
\label{sec:methods}
\subsection{Subjects}
Four healthy subjects without a known history of neuromuscular or orthopedic problems (3 males, 1 female, age 32    $\pm$5, height 177    $\pm$6 cm, mass 75    $\pm$3 kg)  and 2 stroke patients (patient 1: male, age 55, height 178 cm, weight 80 kg, left body-side hemiparesis; patient 2: male, age 35, height 183 cm, weight 80 kg, right body-side hemiparesis; both 3 months after stroke) participated in the study. We have included the patients to introduce the variety in the collected data as they might exhibit different gait to the healthy subjects. Healthy subjects were recruited according to convenience by sampling from the laboratory staff. All participants signed informed consent. This study was approved by the Republic of Slovenia national medical ethics committee and the University rehabilitation institute, Republic of Slovenia ethics committee (80/03/15).

\subsection{Instrumentation}
The experimental setup consisted of a balance assessment robot with an instrumented treadmill (BART) and a set of 8 wireless IMUs. The BART device consists of the instrumented treadmill which measures GRF and COP. Balance assessment robot is a 6 DOF parallel robot which interfaces with the pelvis of the subject. Five of the DOFs (translation of the pelvis in the sagittal, lateral, and vertical directions; pelvic rotation; and pelvic list) are actuated and admittance controlled, while the sixth DOF (pelvic tilt) is passive. {Balance assessment robot uses pelvis brace (PB) to tightly embrace the user in pelvis. All subjects wore special belt that accommodates shape of each subject’s pelvis and in this way ensures that subjects were evenly fastened around waist within PB. This guaranties safety of each subject and avoids fall threatening situations.} A more detailed description of the BART robot can be found in \cite{Olensek2016, Matjacic2019}. The robotic device is capable of delivering perturbations in the forward/backward and left/right directions. In this study, we only considered left/right perturbations delivered in the frontal plane. Perturbations in mediolateral direction during walking are considered more challenging than in the anteroposterior direction and active postural control is needed to regulate mediolateral balance \cite{BAUBY20001433}.

\begin{figure}[h!]
	\centering
	\includegraphics[width=0.5\columnwidth]{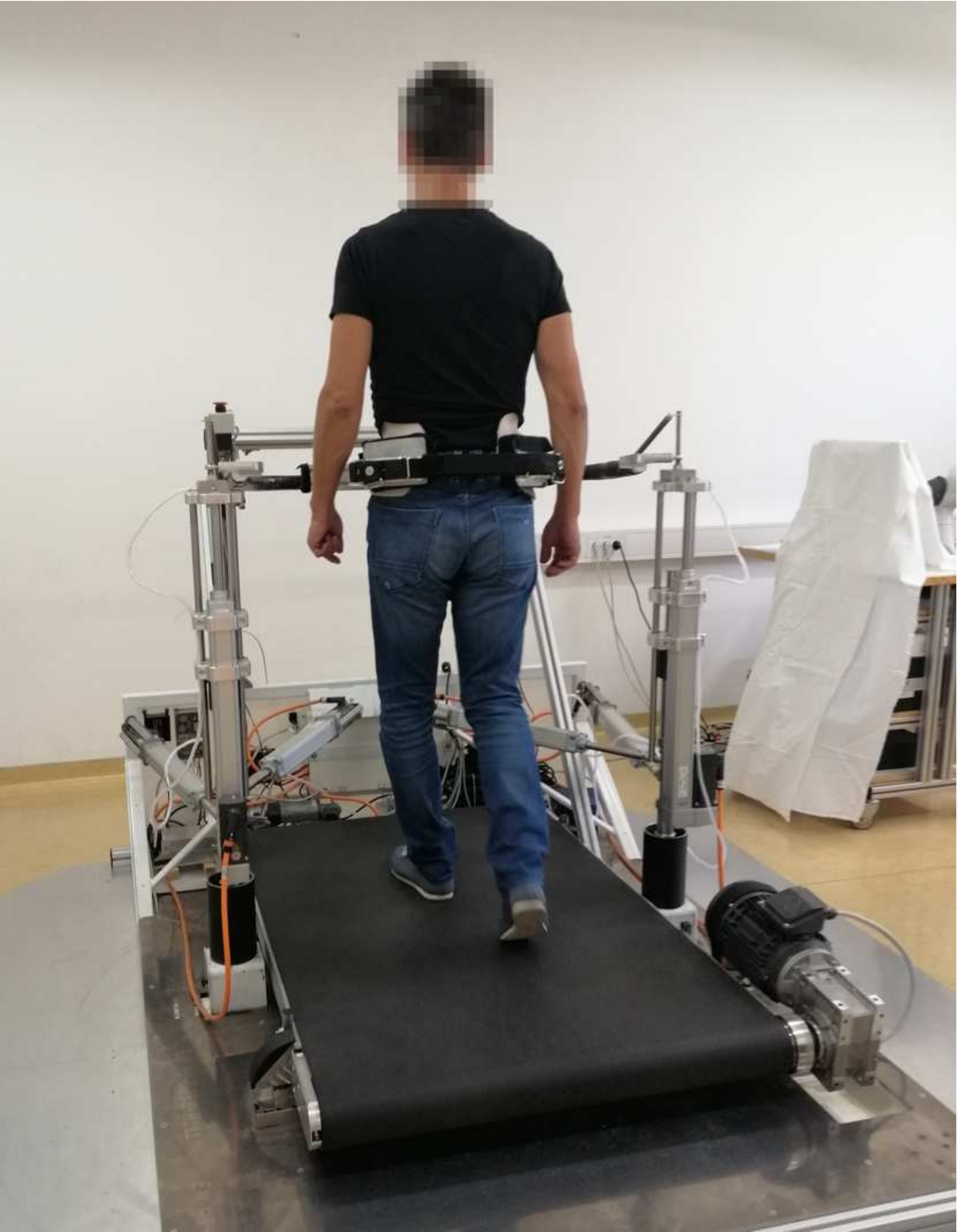}
	\caption{A healthy individual walking on an instrumented treadmill while being embraced by the Balance Assessment Robot for Treadmill walking - BART, which enables perturbations to the subject's pelvis.}
	\label{fig:BART}
\end{figure}

IMUs were developed in our laboratory for the real-time wireless acquisition of the output of the units. Each unit consists of an InvenSense MPU-9250 MEMS device combining a triaxial gyroscope, a triaxial accelerometer, and a triaxial magnetometer in one module, includes a 32-bit Microprocessor Cortex-M4 central processing unit, DWM1000 RF module based on MAC layer of IEEE 802.15.4-2011 (Ultra-wide Band), and a LiPo battery. Each IMU transmits raw accelerometer, gyroscope and magnetometer outputs wirelessly to a receiver unit at a frequency of 100 Hz with the worst-case transmission delay of 0.5 ms. A more detailed description of the IMU device is given in \cite{Logar2019}. IMU system used in this study has the same or better characteristics as a similar system used and validated in \cite{NOVAK20131713, Novak2014}. Both systems, the BART device and the IMUs, were synchronised using an external trigger that aligned the start and the end of data recording.

\subsection{Experimental protocol}
\begin{figure}[h!]
	\centering
	\includegraphics[width=0.5\columnwidth]{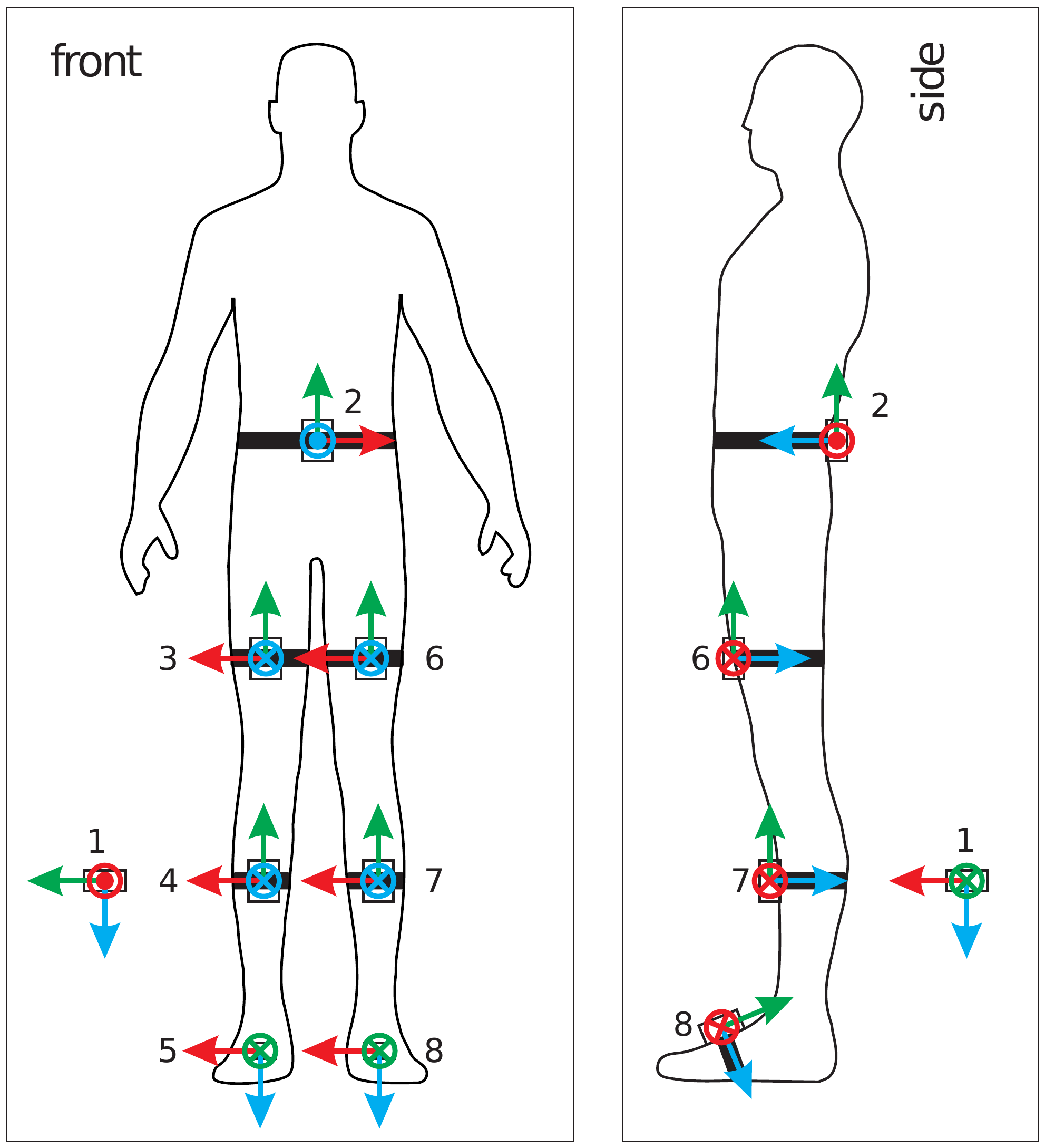}
	\caption{Attachment of wearable IMU sensors. Red axis arrows represent x axis, green axis arrow represent y axis and blue axis arrow represent z axis. All coordinate systems are right handed.}
	\label{fig:IMUsetup}
\end{figure}

COP and GRF were measured using the BART device. The BART device also measured the position and orientation of the pelvis. Lower limb and pelvis movements were measured with 8 wireless IMU units. IMU sensors number 2 to 8 were attached to the pelvis, each thigh, shank, and foot. IMU unit number 1 was positioned far from ferromagnetic materials and the BART device.
Figure \ref{fig:IMUsetup} shows the configuration of IMU units. Coordinate frame of the IMU unit 1 represents the earth reference frame, while IMU unit 2 represents the human body reference frame. Posture in Figure \ref{fig:IMUsetup} defines the initial posture.  

After each subject was properly fitted with IMUs, the subject stepped on the treadmill and was fixed into the BART device using special pelvis belt (PB) that accommodates the shape according to anthropomorphic characteristics of each subject’s pelvis. This ensured that each subject was evenly fastened around the waist by the PB while also preventing relative movement between the subject’s pelvis and the PB. All subjects were familiar with walking in the BART device and had previously walked in the BART device for at least 10 minutes.

Each experiment began with 30 seconds of quiet standing (see Table \ref{tab:ProtocolHealthy}, experiment step 1). This allowed the collection of baseline data. Afterwards, the subject continued with walking experiments. Walking experiments differed by speed and level of perturbations. {In this study perturbations were delivered in mediolateral direction (left/right direction) of the subject}. Perturbations that were used were: no perturbation, perturbation in left/right direction with perturbation force of 10\% of body weight (BW) and perturbation in left/right direction with perturbation force of 15\% of body weight. Perturbation duration was 150 $ms$. As an example, for subject weighting 75 kg, this resulted in 75 $N$ and 112 $N$ of perturbations. {The percentage values used for experiments were based on the results of the previous studies using the BART device \cite{Matjacic2019}. Two levels of perturbations were used, since they elicit different behavior: (a) perturbation was fully contained during the “in-stance” period for perturbation intensities of 10\%, while (b) following a perturbation intensity of 15\%, there was medial displacement of COP in the “stepping period” (from approx. 50\% to approx. 100\% of a gait cycle) that finally contained the instability.} There was at least six seconds recovery period between two perturbations that allowed each subject to fully recover from perturbation. Throughout the experiment, each subject was given visual feedback on a laptop screen that in real-time graphically illustrated current pelvis position with respect to the center of available range. When selecting perturbations parameters the goal was to select such perturbation amplitude that would elicit substantial balancing responses while not creating fall threatening situations. Full experimental protocol is given in Table \ref{tab:ProtocolHealthy}. Experiments concluded with quiet standing for 30 seconds (see Table \ref{tab:ProtocolHealthy}, experiment step 9). After each walking experiment step subjects had 1 minute of resting period.

\begin{table}[h]
	\caption{Measurement protocol for healthy subjects. BW - body weight, QS - quiet standing.}	
	\label{tab:ProtocolHealthy}
	\centering
	\begin{tabular}{lccc} 
		\toprule
		exp. step & walking speed $[\frac{\text{m}}{\text{s}}]$ & perturbation $[\%$ of BW$]$ & duration $[\text{s}]$ \\ [0.5ex] 
		\cmidrule(r){2-4}
		1 & 0 (QS) & 0 & 30\\ 
		2 & 0.38 & 0 & 180\\
		3 & 0.38 & 10 & 420\\
		4 & 0.38 & 15 & 420\\
		5 & 0.5 & 0 & 180\\		
		6 & 0.5 & 10 & 420\\ 
		7 & 0.5 & 15 & 420\\ 
		8 & 0.8 & 0 & 180\\ 
		9 & 0 (QS) & 0 & 30 \\
		\bottomrule
	\end{tabular} \vspace{3pt}
\end{table}

For stroke subjects the protocol was adapted: experiments with 15 \% level of perturbation and walking speed of 0.8 $\frac{m}{s}$ were omitted. This was done for safety reasons and to prevent the stroke subject from getting overly fatigued. Full measurement protocol for stroke subjects is given in Table \ref{tab:ProtocolStroke}.

\begin{table}[h]
	\caption{Measurement protocol for stroke subjects. BW - body weight, QS - quiet standing.}
	\centering
	\begin{tabular}{lccc} 
		\toprule
		exp. step & walking speed $[\frac{\text{m}}{\text{s}}]$ & perturbation $[\%$ of BW$]$ & duration $[\text{s}]$ \\ [0.5ex] 
		\cmidrule(r){2-4}
		1 & 0 (QS) & 0 & 30\\ 
		2 & 0.38 & 0 & 180\\
		3 & 0.38 & 10 & 420\\
		4 & 0.5 & 0 & 180\\
		5 & 0.5 & 10 & 420\\ 
		6 & 0 (QS) & 0 & 30 \\
		\bottomrule
	\end{tabular} \vspace{3pt}
	\label{tab:ProtocolStroke}
\end{table}

\subsection{Data}

\paragraph{Our dataset}
The data we recorded for each of the six subjects consists of:
\begin{itemize}
	\item the Cartesian components of gyroscope, accelerometer, and magnetometer (GAM) raw signals from 8 IMUs measured in each IMU frame,
	\item the COP $x,y$ location in the BART frame,
	\item the centre of subjects' pelvis $x, y$ coordinates in the BART frame,
	\item and the IMU-BART synchronization signal.
\end{itemize}
We were interested in the COP coordinates relative to each subject's position rather than absolute coordinates in the BART frame. Therefore, we transformed the COP coordinates measured in the BART frame into the centre of pelvis frame\footnote{Note that the centre of the pelvis is very close to the COM of a human body during walking \cite{COMPELVIS1}.}.

The error in IMU signals due to random fluctuations is about 3\%. This was estimated by comparing the size of random fluctuations recorded by the static IMU, which was not attached to either the subjects or BART, with the size of the signals on IMUs attached to the walking subjects. 

Measurement characteristics of the BART device are covered in \cite{Olensek2016, Matjacic2019}.
\paragraph{Public dataset}
In addition to our dataset, we also considered a public dataset \cite{synthetic}, which recorded 42 healthy subjects during walking. Their kinematics was measured using markers and a three-dimensional motion-capture system. The forces during walking were collected using an instrumented treadmill, measuring the GRF as well as the COP. The experimental protocol consisted of eight trials of different walking speeds and a single static trial, in total reaching about 4 minutes of recordings per subject. We consider 33 subjects that completed the full protocol.

\begin{figure}[h!]
	\centering
	\includegraphics[width=1.0\columnwidth]{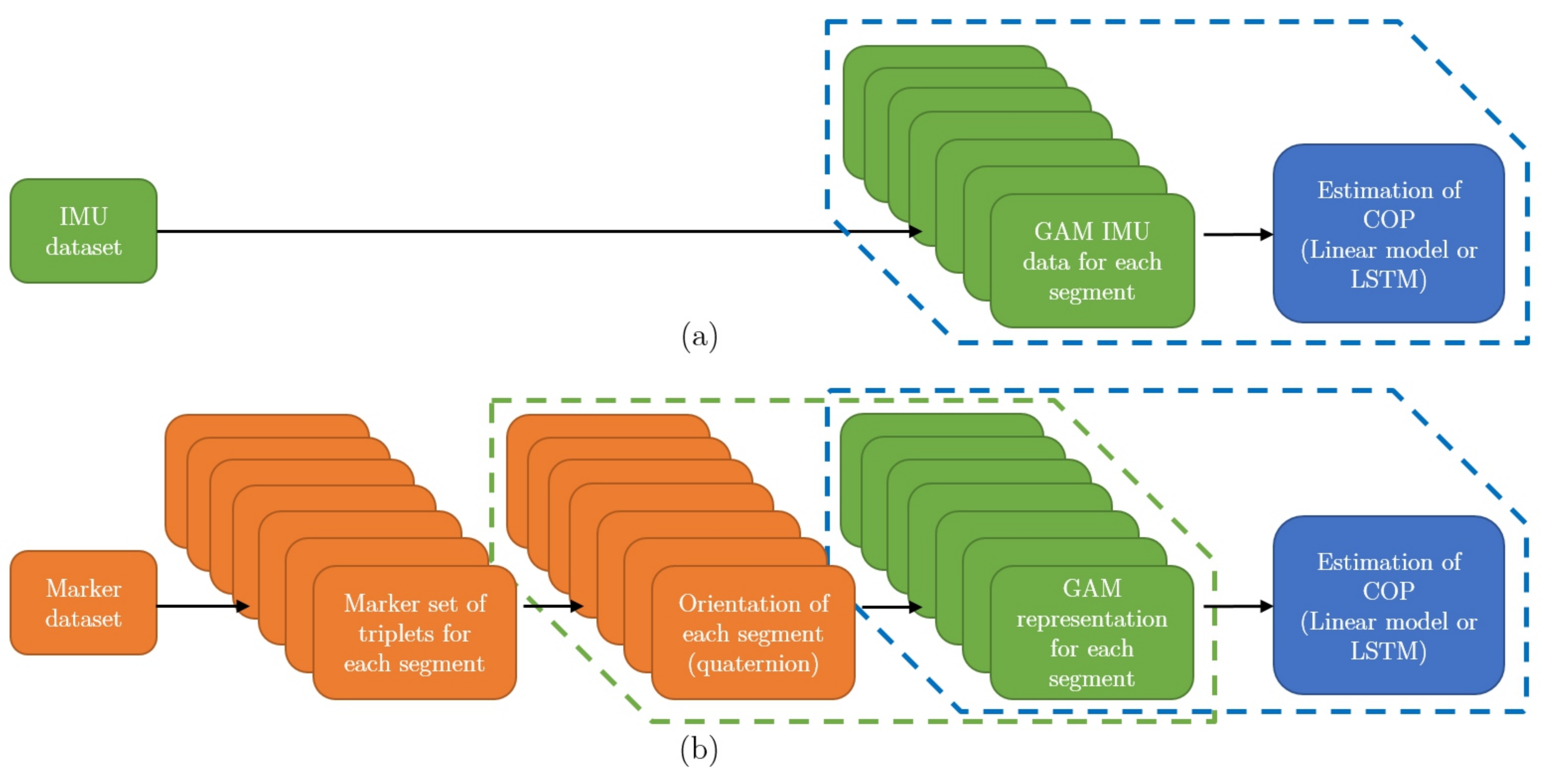}
	\caption{{Figure shows an algorithm steps used for estimating COP from public dataset. Subfigure (a) shows the steps for estimating COP from IMU data. IMU dataset is divided into data consisting of a gyroscope, accelerometer, and magnetometer (GAM) raw signals for each segment. This data is then fed into algorithm for estimating COP. Subfigure (a) is added to this figure as a comparison to show which are additional steps required for estimating COP from public dataset which are shown in subfigure (b). Marker data from public dataset are divided into marker triplets attached to particular human body segments and are used to calculate the orientation of each segment. Orientation of each segment is represented in quaternion form. In next step, a quaternion form representation of orientation of each segment is converted into GAM form. This form allows the use of same algorithms for estimating the COP as in case of using IMU data. A more detailed block diagram for steps framed by green dashed line are shown in Figure \ref{fig:workflow2}. A more detailed block diagram for steps framed by blue dashed line for estimating COP from GAM data (IMU or GAM representation for public dataset) is shown in Figure \ref{fig:lstm_diag}.}}
	\label{fig:workflow1}
\end{figure}

\begin{figure}[h!]
	\centering
	\includegraphics[width=0.8\columnwidth]{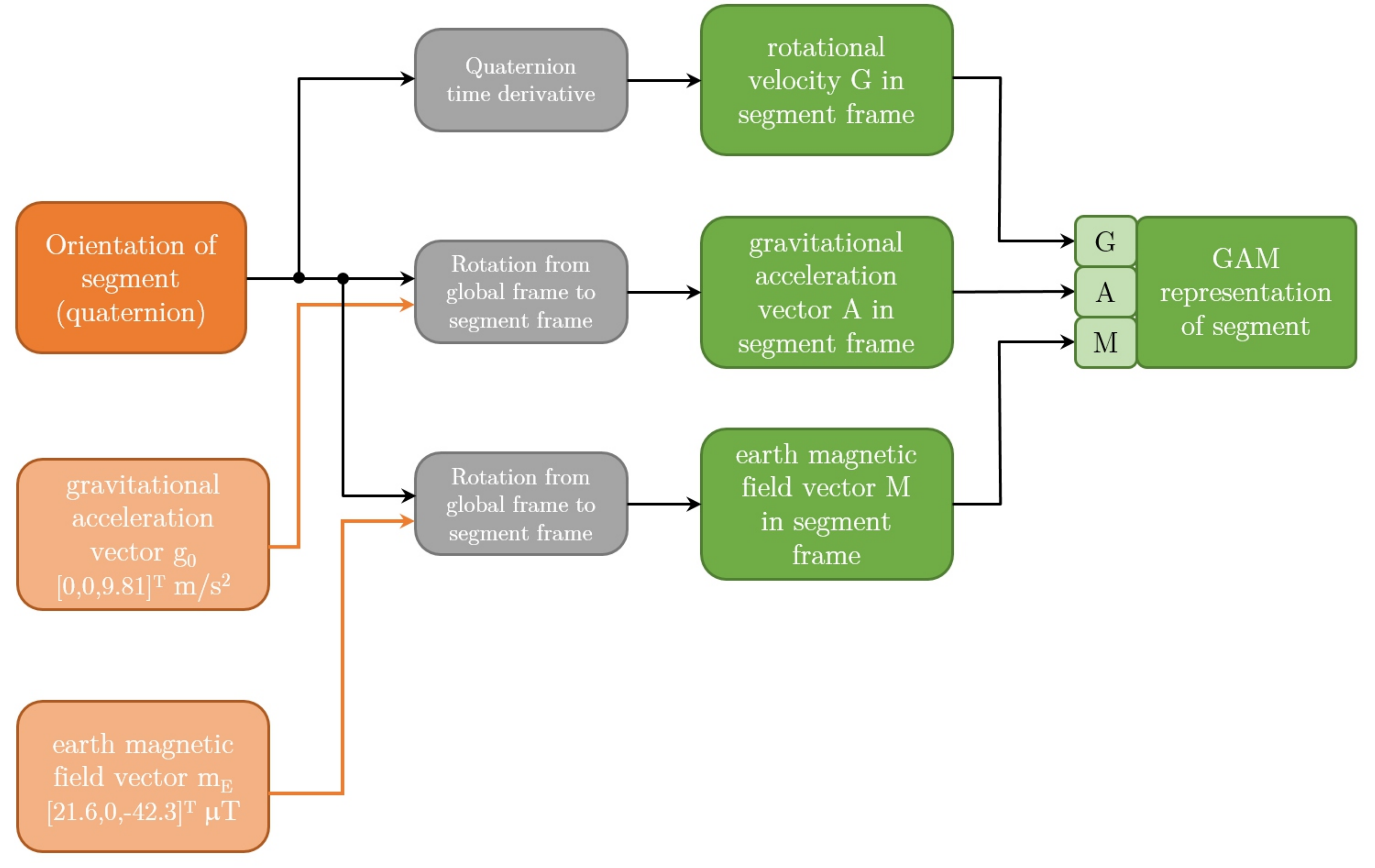}
	\caption{{Figure shows the calculation of GAM representation from quaternion representation.}}
	\label{fig:workflow2}
\end{figure}

Using the marker data we calculated the orientation of each segment following the methodology presented in \cite{CAPPOZZO1995171} in quaternion representation. {To calculate orientation of the segment from marker positions three markers are required for each segment. Marker dataset is therefore divided in triplets of markers for each segment. The steps of algorithm are shown in Figures \ref{fig:workflow1} and explained in caption of the Figure \ref{fig:workflow1}}. Next, the quaternions were rotated by fixed rotation so that coordinate frame attached to the segment match relative orientation of the coordinates frames of the IMUs used in our experiment (see Fig. \ref{fig:IMUsetup}). Next, we represented the orientations in the form of GAM triplets. {The algorithm for converting representation from quaternion form into GAM representation is shown in Figure \ref{fig:workflow2}}. G component is rotational velocity which is calculated by differentiation of the quaternion of the segment \cite{graf2008quaternions}, while A and M components are calculated by rotating the earth gravitational and magnetic field vectors by quaternion representing the orientation of the segment. While GAM representation is not common representation, it does not lose any information about the orientation of the segment, since the quaternion representing the orientation of the segment can be calculated from A and M components using \textsc{quest} algorithm \cite{quest}. We also translated the measured COP in the treadmill frame to the centre of pelvis frame. This way we restructured the data in the public dataset into the same form as our own measurements, that is, GAM values for all three axis with the corresponding COP.

\subsection{The IMU-COP mapping, aims, and model choices}
\label{sec:maping_aims}
In this paper we focus on the \emph{end-to-end} mapping between the raw IMU signal in the form of GAM values (the input) and the COP position (the output) during walking. Therefore, we do not apply any preprocessing of the IMU signals or manually construct deterministic models of the kinematics and dynamics of human walking. We feed the IMU data directly into statistical models that in turn return the COP coordinates. The models are trained by minimizing the mean squared error (MSE) between the predicted and the measured COP using a part of the recorded data (`train' data). The predictive performance of models is then evaluated in terms of RMS error on the unseen (`test') data.

The COP coordinates can in principle be calculated using the measured IMU signal (after de-biasing and de-noising) using a physical model of the human body, an estimation of body segment masses and moments of inertia, orientations, forces, and a series of coordinate transformations. In statistical modelling, all of this is encoded in parameters or weights of models which are learned empirically from data, resulting in a simpler end-to-end modelling process.

Our guiding principle is modelling simplicity and ease of use of the IMU wearable measurement system. Consequently, we address the following questions:
\begin{enumerate}
	\item How does the predictive performance of models depend on the amount of training data used?
	
	We would like to minimise the time subjects or patients need to spend walking in the BART device generating train set data.
	
	\item How does the predictive performance depend on the number \emph{and} placement of IMUs?
	
	We would like to find the minimal configuration and optimal placement of IMUs that the subjects need to wear while ensuring that the predictive performance of models does not degrade significantly.
	
	\item Which type of signal can be discarded without degradation in predictive performance?
	
	We would like to remove some of the magneto-inertial data recorded by IMUs. For example, the magnetic field measurement is very sensitive to changes in the environment, thus removing this data from the model input makes the wearable system more robust.
	
	\item How transferable are the subject-specific models to other subjects?
	
	We would like to determine how well the models trained on one set of subjects generalises to the unseen data of another set of subjects. The aim is to use the wearable system on new subjects without the extra measurements and calibration in the BART device.
\end{enumerate}

\paragraph{Model choices and training} 
Many statistical models can be used to approximate the IMU-COP mapping: linear models, decision trees, random forests, support vector machines, ANNs of different architectures, and many more.

We choose a linear model as a baseline for modelling the IMU-COP mapping and select a Long Short-Term Memory \cite{LSTMorigin} artificial neural network (LSTM) out of many possible non-linear models{\footnote{We implement the models using the \textit{TensorFlow}\cite{tensorflow2015-whitepaper} and \textit{scikit-learn}\cite{scikit-learn} libraries in \textit{python}.}. LSTM neural networks consist of units (neurons) that are capable of storing sequential values, giving them, for example, the ability to construct integrals and derivatives of input data, which makes them widely used and well suited for time series modelling. For a review of LSTM neural networks see \cite{RNNreview}. The linear model we use is a standard multivariate linear regression of input IMU data on the output COP data, where mean squared error is minimized.}

\begin{figure}[h!]
	\centering
	\includegraphics[width=0.75\columnwidth]{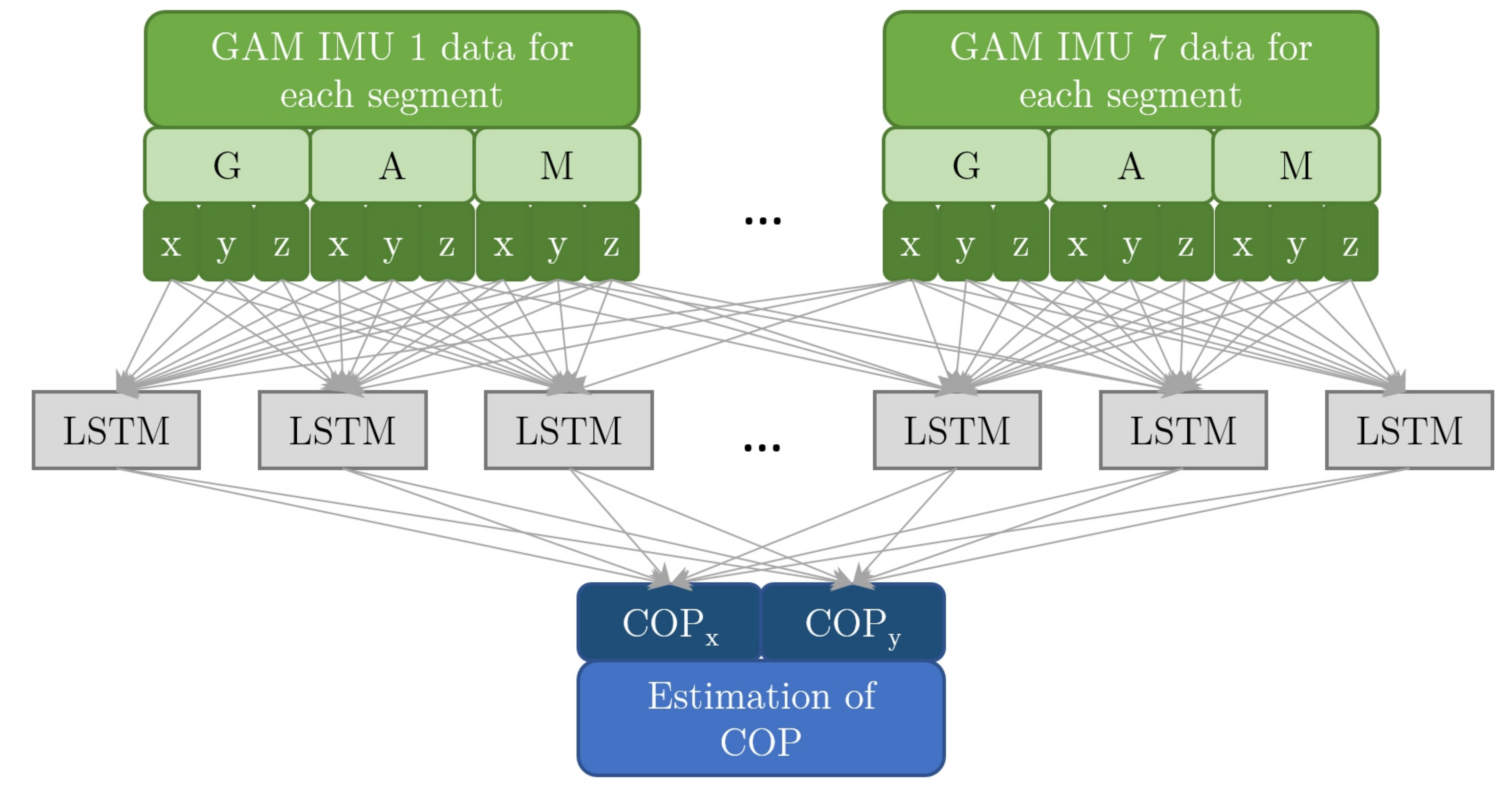}
	\caption{Architecture of the IMU-COP map using LSTM artificial neural network.}
	\label{fig:lstm_diag}
\end{figure}

The LSTM neural network we use consists of a single `hidden' layer of 100 LSTM units. For the diagram of the modelling architecture see Fig.\,\ref{fig:lstm_diag}. We have checked the sizes ranging from 10 to 1000 hidden units and found that above roughly 100 units there was no significant improvement in the predictive performance. Training of the LSTM network was done with the ADAM optimizer \cite{ADAM} minimizing the mean squared error, with learning rate 0.001 until train error stopped decreasing. The length of memory for the LSTM was limited to the most recent 0.1\,s as there was no improvement in predictive performance with longer LSTM memory. We do not consider more elaborate models because we do not focus on finding the best network architectures, choice of hyperparameters, or training methods, as explained in Sec.\,\ref{sec:maping_aims}. 

\section{Results and Discussion}
\label{sec:results}
\begin{figure}[h!]
    \begin{subfigure}[b]{0.5\textwidth}
    		\centering
    		\includegraphics[width=\columnwidth]{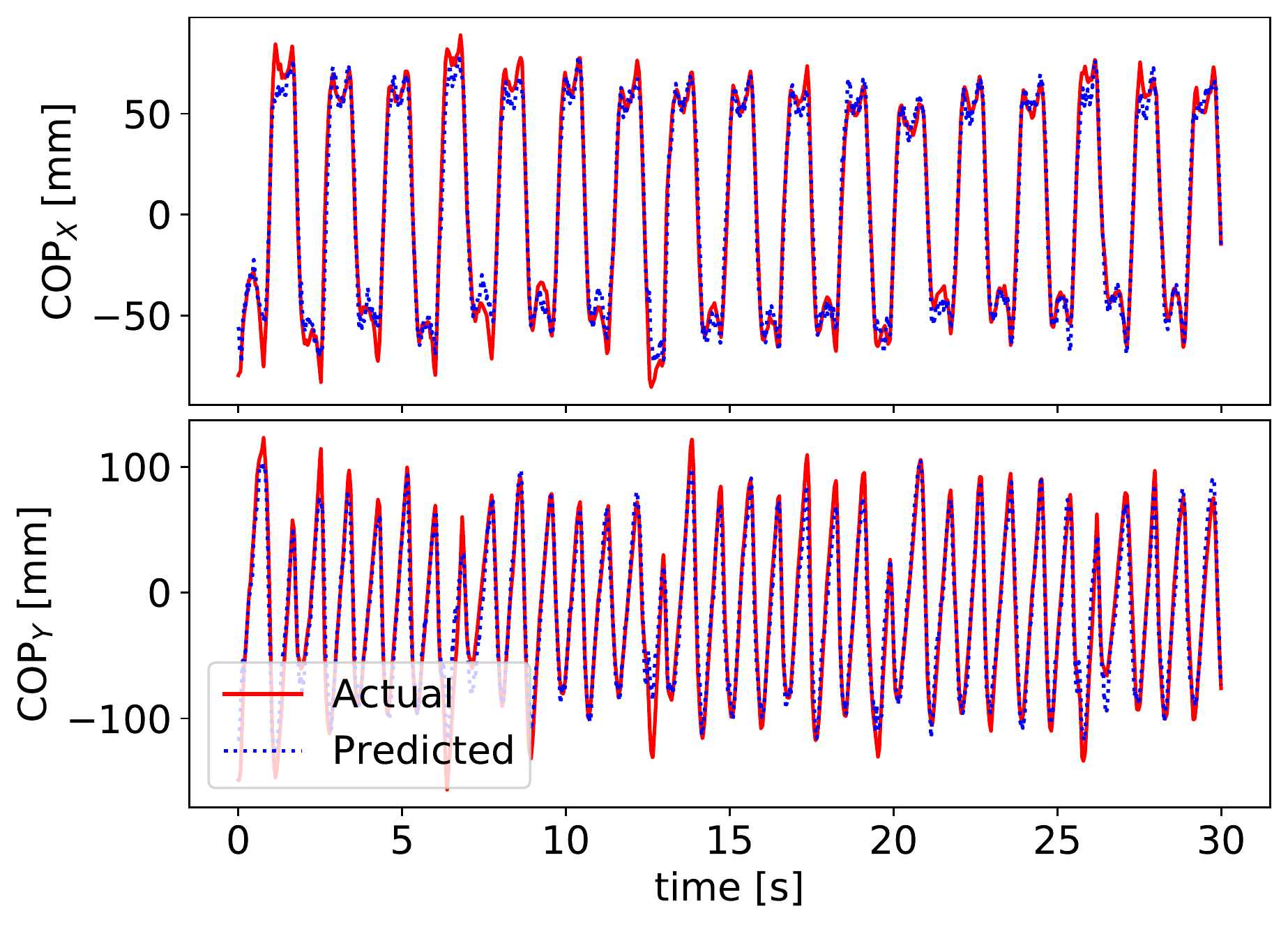}
    		\caption{}
    		\label{fig:lstm_pred}
    	\end{subfigure}
    	\begin{subfigure}[b]{0.5\textwidth}
    		\centering
    		\includegraphics[width=\columnwidth]{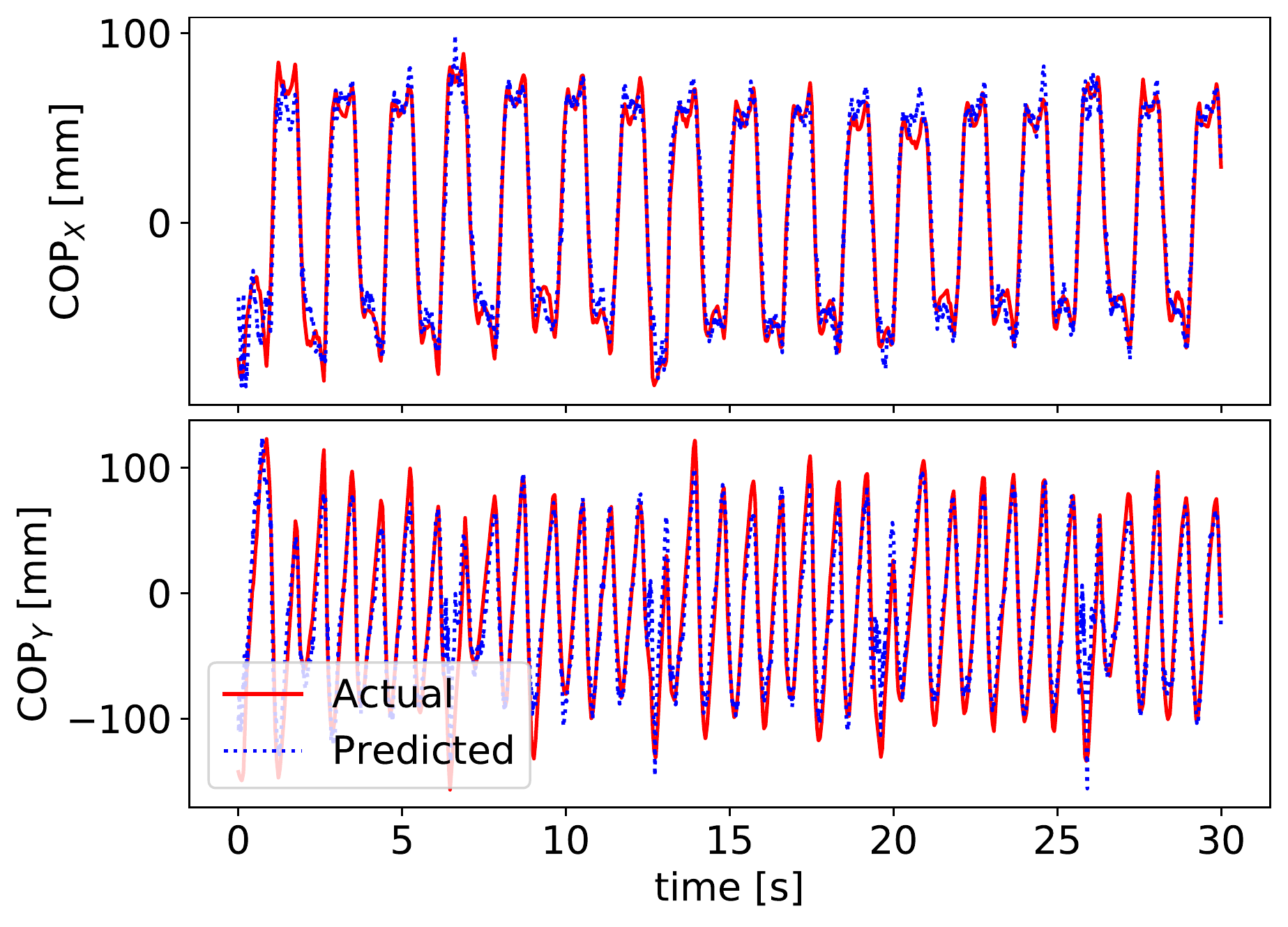}
    		\caption{}
    		\label{fig:lin_pred}
    	\end{subfigure}
    	\vskip\baselineskip
    	\hspace{0.25\textwidth}%
    	\begin{subfigure}[b]{0.5\textwidth}
    		\centering
    		\includegraphics[width=\columnwidth]{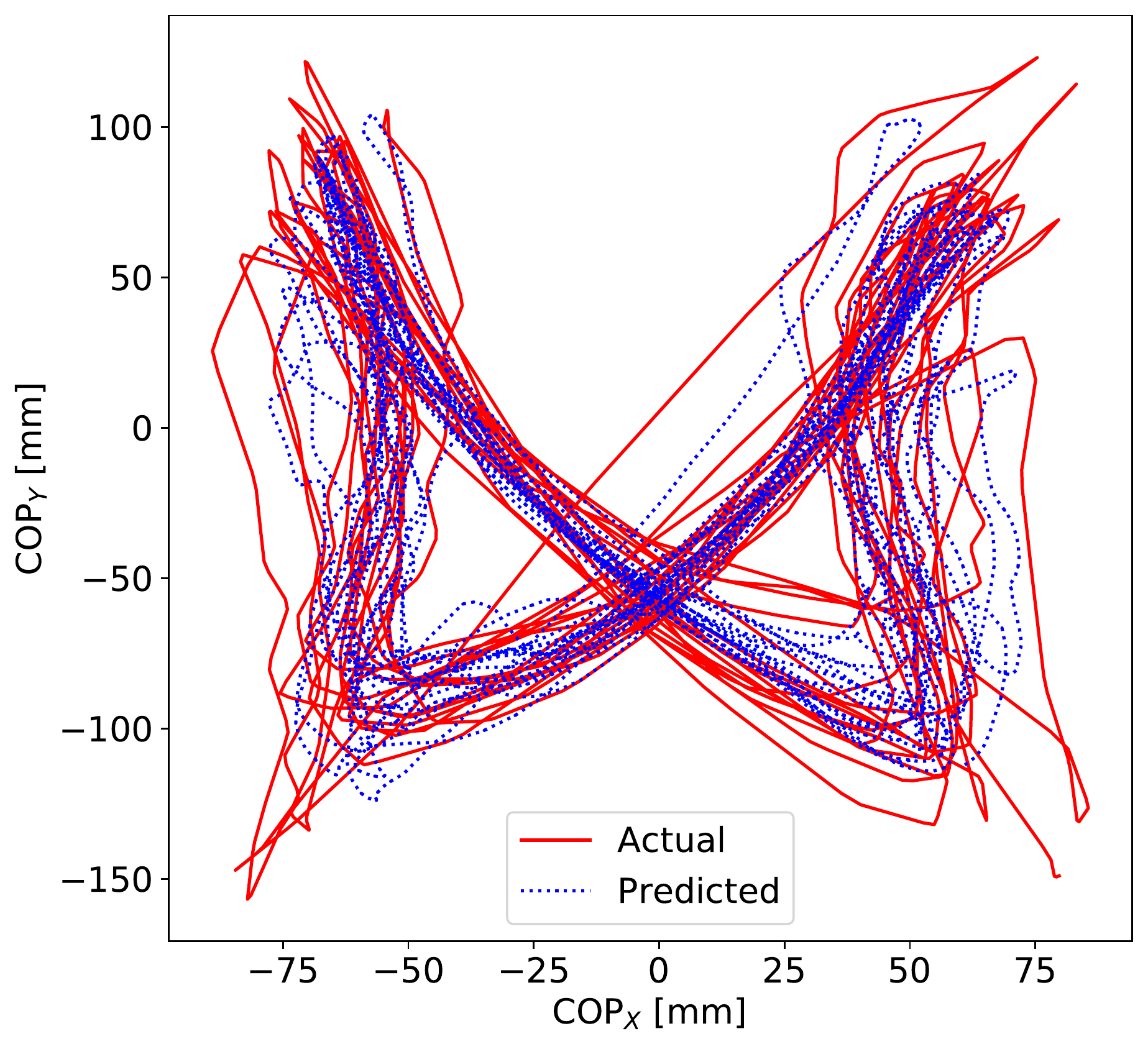}
    		\caption{}
    		\label{fig:lstm_trans}
    	\end{subfigure}

	\caption{Prediction and ground truth for (COP$_x$, COP$_y$) for subject S3 for a sample of 30\,s. (\textbf{a}) LSTM prediction for subject S3 trained on S3 data. (\textbf{b}) Linear model prediction for subject S3 trained on S3 data.
	(\textbf{c}) LSTM prediction for subject S3 trained on S3 data.} \label{fig:COP}
\end{figure}

Signals for the ground truth COP, measured by the BART device, and the estimated COP are shown in Figure \ref{fig:COP}. 

\begin{figure}[h!]
	\centering
	\includegraphics[width=\columnwidth]{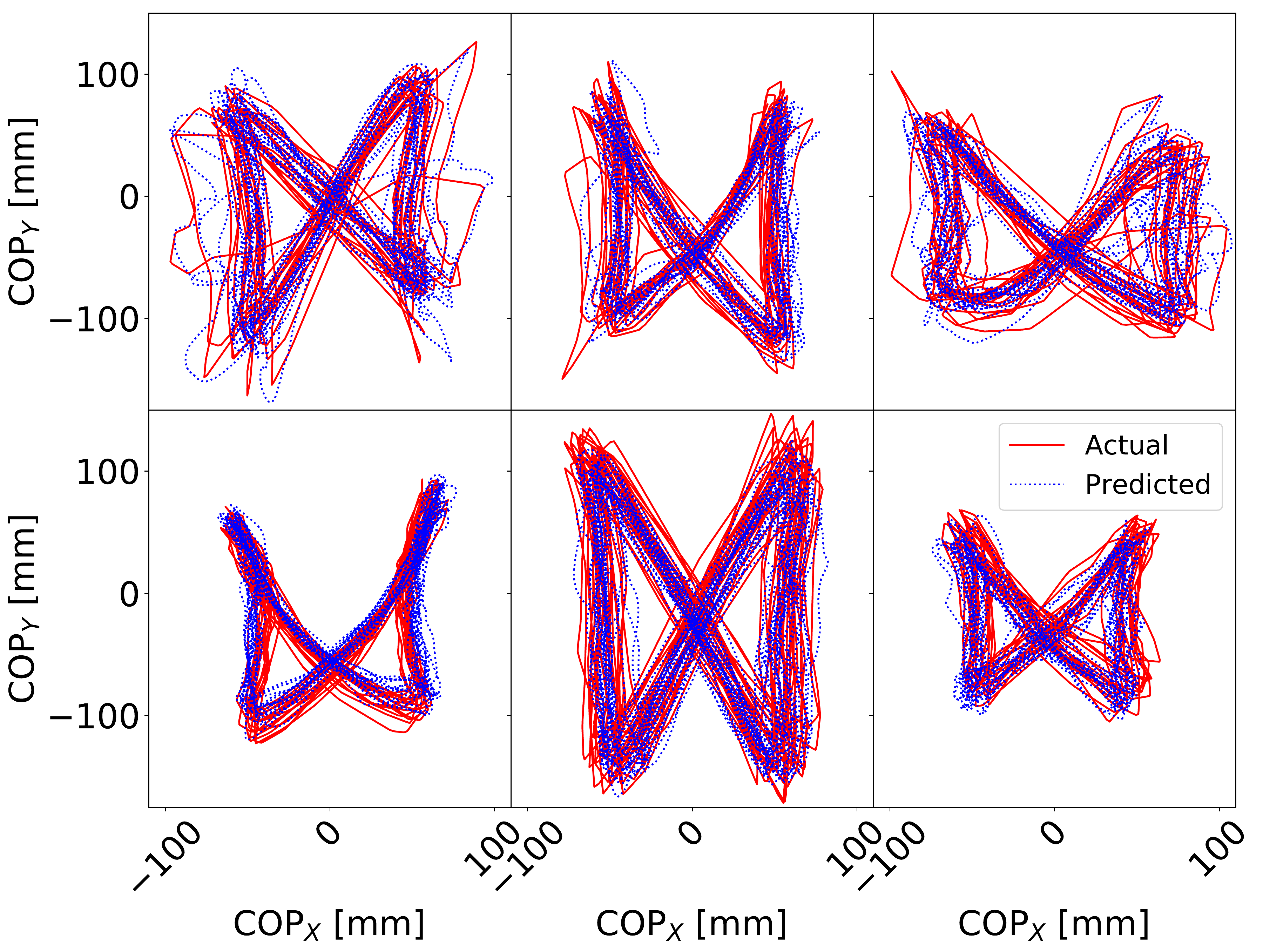}
	\caption{Gaitograms for the subjects measured in this study. Top left - S1, top middle - S2, bottom left - S3, bottom middle- S4, top right - P1, bottom right - P2.}
	\label{fig:gaitograms}
\end{figure}

Figures \ref{fig:COP} (a) and (b) show signals of COP$_x$ (anterior) and COP$_y$ (lateral) for subject S3. Figure \ref{fig:COP} (c) shows x-y plot of COP signals. Figure \ref{fig:COP} (c) shows a typical example of the "butterfly" pattern of the COP \cite{ROERDINK20082628, VANDERVEEN2018218} in the centre of the pelvis frame. This characteristic shape of the COP pattern is called a gaitogram and gives important information about specific gait events and gait characteristics: fore-after symmetry gives information about anterior-posterior foot displacements during the single-support stance phase, while left-right symmetry gives information about weight bearing during single-support stance phase (for more in-depth description and analysis of gaitogram please see \cite{ROERDINK20082628, zadrnovel, Spelitz2020}). Figure \ref{fig:gaitograms} shows examples of gaitograms for all test subjects.

\begin{table*}[h]
	\centering
	\caption{Intra-subject test RMS error [mm] for various input data. The total RMS error (Tot.) is split between the lateral (Lat.) and anterior (Ant.) components. Columns `GAM' label models with the full input of gyroscope, accelerometer, and magnetometer data from all IMUs. Column `GAM+hist' additionally includes the most recent 0.1\,s IMU data in the input. Note,  that the inclusion of `history' in the input to linear model results in significant improvement in the RMS error.}
	\label{tab:intra}
	\begin{tabular}{lcccccccccccccccc}
		\toprule
		\multicolumn{1}{c}{} & \multicolumn{9}{c}{Linear model} &\multicolumn{6}{c}{LSTM} \\ 
		\cmidrule(r){2-10} \cmidrule(r){11-16} 
		& \multicolumn{3}{c}{GAM} & \multicolumn{3}{c}{GAM+hist} & \multicolumn{3}{c}{GA}  & \multicolumn{3}{c}{GAM} & \multicolumn{3}{c}{GA}  \\
		\cmidrule(r){2-4} \cmidrule(r){5-7} \cmidrule(r){8-10} \cmidrule(r){11-13} \cmidrule(r){14-16}
		Subj.       & Tot. & Lat. & Ant.    & Tot. & Lat. & Ant.    & Tot. & Lat. & Ant.    & Tot. & Lat. & Ant.   & Tot. & Lat. & Ant.  \\
		S1          & 21.8 & 17.3 & 25.5    & 18.8 & 15.2 & 21.9    & 23.8 & 18.3 & 28.3    & 13.7 & 9.2 & 17.0    & 14.4 & 10.1 & 17.8  \\
		S2          & 21.0 & 15.6 & 25.3    & 18.4 & 13.4 & 22.3    & 24.7 & 17.2 & 30.4    & 12.8 & 8.8 & 15.8    & 13.9 & 9.7 & 17.1   \\
		S3          & 17.4 & 12.1 & 21.5    & 14.6 & 10.4 & 17.8    & 19.8 & 12.9 & 24.9    & 10.3 & 7.2 & 12.7    & 11.0 & 7.5 & 13.6  \\
		S4          & 18.5 & 14.3 & 21.8    & 15.9 & 12.4 & 18.7    & 21.7 & 15.2 & 26.6    & 11.3 & 8.1 & 13.8    & 11.7 & 8.6 & 14.1  \\
		P1          & 19.0 & 16.5 & 21.2    & 16.9 & 14.7 & 18.8    & 21.2 & 17.6 & 24.3    & 12.5 & 10.5 & 14.3   & 14.0 & 11.5 & 16.0  \\
		P2          & 17.2 & 13.2 & 20.4    & 15.3 & 11.9 & 18.2    & 18.7 & 14.7 & 22.0    & 13.4 & 10.2 & 16.0   & 14.2 & 11.0 & 16.9  \\
		\cmidrule(r){2-4} \cmidrule(r){5-7} \cmidrule(r){8-10} \cmidrule(r){11-13} \cmidrule(r){14-16}
		AVG        & 19.2 & 14.8 & 22.6 	& 16.7 & 13.0 & 19.6 	& 21.7 & 16.0 & 26.1    & 12.3 & 9.0 & 14.9   & 13.2 & 9.7 & 15.9  \\
		\bottomrule
	\end{tabular}
\end{table*}

\begin{table*}[h!]
	\centering
	\caption{Intra-subject test RMS error [mm] for the public dataset \cite{synthetic}. The total RMS error (Tot.) is split between the lateral (Lat.) and anterior (Ant.) components. Columns `GAM' label models with the full input of gyroscope, accelerometer, and magnetometer data from all IMUs. Column `GAM+hist' additionally includes the most recent 0.1\,s IMU data in the input.}
	\label{tab:intra_synt}
	\begin{tabular}{lcccccccccccccccc}
		\toprule
		\multicolumn{1}{c}{} & \multicolumn{9}{c}{Linear model} &\multicolumn{6}{c}{LSTM} \\ 
		\cmidrule(r){2-10} \cmidrule(r){11-16} 
		& \multicolumn{3}{c}{GAM} & \multicolumn{3}{c}{GAM+hist} & \multicolumn{3}{c}{GA}  & \multicolumn{3}{c}{GAM} & \multicolumn{3}{c}{GA}  \\
		\cmidrule(r){2-4} \cmidrule(r){5-7} \cmidrule(r){8-10} \cmidrule(r){11-13} \cmidrule(r){14-16}
		Subj.       & Tot. & Lat. & Ant.    & Tot. & Lat. & Ant.    & Tot. & Lat. & Ant.    & Tot. & Lat. & Ant.   & Tot. & Lat. & Ant.  \\
        01      &        17.2 &          12.2 &           21.0 &        13.5 &           9.3 &           16.6 &        17.7 &          12.4 &           21.7 &        10.6 &           5.8 &           13.8 &        10.8 &           6.2 &           14.0 \\
02      &        17.7 &          11.8 &           22.1 &        13.5 &           9.6 &           16.4 &        18.0 &          11.8 &           22.5 &        10.7 &           6.3 &           13.7 &        11.3 &           7.2 &           14.2 \\
03      &        15.3 &           9.3 &           19.6 &        11.9 &           7.4 &           15.2 &        15.6 &           9.5 &           19.9 &         9.2 &           4.9 &           12.1 &         9.2 &           5.3 &           11.8 \\
04      &        19.7 &          12.2 &           25.0 &        15.5 &          10.6 &           19.2 &        19.9 &          12.3 &           25.3 &        12.0 &           6.5 &           15.8 &        12.2 &           7.0 &           15.9 \\
05      &        16.5 &          10.4 &           20.8 &        13.1 &           8.2 &           16.6 &        16.8 &          10.7 &           21.2 &        10.0 &           5.2 &           13.2 &        10.4 &           5.7 &           13.5 \\
06      &        15.7 &          11.5 &           19.1 &        11.7 &           8.3 &           14.3 &        17.0 &          12.2 &           20.6 &         8.9 &           5.6 &           11.2 &         9.0 &           5.7 &           11.4 \\
07      &        15.7 &          11.8 &           18.7 &        13.5 &          10.0 &           16.2 &        16.9 &          11.8 &           20.7 &         9.6 &           5.7 &           12.3 &        10.0 &           6.6 &           12.6 \\
08      &        18.3 &          12.5 &           22.7 &        13.6 &          10.3 &           16.3 &        18.4 &          12.5 &           22.8 &        10.4 &           5.9 &           13.4 &        10.9 &           6.5 &           14.0 \\
09      &        16.7 &          11.9 &           20.5 &        13.5 &           9.8 &           16.3 &        17.2 &          12.0 &           21.1 &         9.7 &           5.9 &           12.4 &        10.4 &           6.2 &           13.3 \\
10      &        14.6 &          11.1 &           17.4 &        12.1 &           8.5 &           14.9 &        16.6 &          11.1 &           20.7 &        10.8 &           5.9 &           14.1 &        11.4 &           7.0 &           14.5 \\
11      &        20.0 &          12.7 &           25.4 &        15.3 &          10.3 &           19.1 &        20.8 &          13.4 &           26.2 &        10.9 &           6.2 &           14.2 &        11.3 &           7.0 &           14.3 \\
12      &        17.6 &          12.4 &           21.6 &        14.6 &          10.3 &           17.8 &        18.1 &          12.6 &           22.4 &        12.3 &           6.7 &           16.0 &        12.6 &           6.9 &           16.4 \\
13      &        14.9 &          10.9 &           18.0 &        11.1 &           9.2 &           12.8 &        15.0 &          10.9 &           18.2 &         8.4 &           5.8 &           10.4 &         8.9 &           6.4 &           10.8 \\
14      &        13.7 &           8.6 &           17.3 &        10.2 &           7.4 &           12.4 &        15.8 &           8.5 &           20.7 &         8.6 &           4.8 &           11.2 &         8.8 &           5.1 &           11.4 \\
15      &        18.7 &          13.6 &           22.6 &        16.1 &          11.1 &           19.9 &        20.6 &          13.8 &           25.6 &        11.7 &           6.9 &           15.0 &        12.6 &           7.7 &           16.1 \\
16      &        16.0 &          11.2 &           19.7 &        12.5 &           8.9 &           15.3 &        16.4 &          11.3 &           20.2 &        10.4 &           5.9 &           13.4 &        11.0 &           6.2 &           14.3 \\
18      &        18.1 &          11.7 &           22.8 &        14.9 &           9.3 &           18.8 &        19.4 &          11.8 &           24.8 &        11.4 &           6.3 &           14.9 &        12.2 &           7.0 &           15.8 \\
19      &        16.3 &           9.3 &           21.1 &        12.5 &           7.8 &           15.9 &        17.0 &           9.6 &           22.0 &        11.1 &           6.3 &           14.4 &        10.5 &           6.0 &           13.6 \\
20      &        13.9 &           8.1 &           17.9 &        11.7 &           6.9 &           15.0 &        14.4 &           8.4 &           18.5 &         9.3 &           5.2 &           12.1 &        10.3 &           5.7 &           13.3 \\
21      &        18.4 &          12.5 &           22.8 &        14.8 &          10.0 &           18.3 &        18.6 &          12.5 &           23.1 &        12.6 &           6.2 &           16.7 &        12.7 &           6.6 &           16.7 \\
22      &        19.5 &          12.9 &           24.4 &        15.3 &          11.0 &           18.6 &        20.7 &          13.1 &           26.2 &        11.9 &           6.8 &           15.3 &        12.6 &           7.5 &           16.1 \\
23      &        16.0 &          11.0 &           19.7 &        12.7 &           8.3 &           15.9 &        16.2 &          11.0 &           20.0 &        10.9 &           6.5 &           14.0 &        10.7 &           6.4 &           13.7 \\
24      &        14.6 &          12.0 &           16.8 &        11.5 &          10.1 &           12.8 &        14.7 &          12.0 &           17.0 &         7.9 &           5.2 &            9.8 &         8.4 &           5.7 &           10.4 \\
25      &        16.9 &          13.5 &           19.8 &        13.2 &          10.5 &           15.4 &        17.5 &          13.9 &           20.6 &         9.1 &           5.7 &           11.5 &         9.8 &           6.5 &           12.3 \\
26      &        16.1 &          13.1 &           18.7 &        12.4 &          10.1 &           14.3 &        16.8 &          13.1 &           19.8 &        10.2 &           6.6 &           12.9 &        10.7 &           7.0 &           13.4 \\
27      &        17.1 &          15.0 &           19.0 &        14.4 &          12.2 &           16.3 &        17.3 &          15.2 &           19.2 &        10.8 &           7.1 &           13.5 &        11.2 &           7.3 &           14.0 \\
30      &        12.9 &           8.6 &           16.0 &        10.9 &           7.4 &           13.5 &        13.6 &           8.7 &           17.2 &         8.6 &           4.7 &           11.3 &         9.0 &           4.9 &           11.7 \\
31      &        13.0 &           9.5 &           15.7 &        10.9 &           7.7 &           13.3 &        13.0 &           9.7 &           15.6 &         8.4 &           5.1 &           10.7 &         8.9 &           5.3 &           11.4 \\
33      &        21.1 &          12.7 &           27.1 &        17.5 &          10.6 &           22.3 &        22.1 &          13.1 &           28.4 &        13.6 &           7.2 &           17.8 &        14.0 &           8.0 &           18.1 \\
34      &        17.7 &          11.7 &           22.1 &        15.1 &          10.3 &           18.8 &        18.0 &          11.6 &           22.7 &        12.4 &           7.3 &           16.0 &        13.0 &           7.4 &           16.8 \\
35      &        19.9 &          13.2 &           24.9 &        16.1 &          10.0 &           20.5 &        20.2 &          13.4 &           25.2 &        13.9 &           6.9 &           18.4 &        14.0 &           7.5 &           18.4 \\
38      &        19.7 &          12.8 &           24.8 &        14.8 &          10.3 &           18.2 &        19.9 &          13.1 &           24.9 &        12.5 &           6.4 &           16.4 &        12.5 &           7.2 &           16.2 \\
40      &        17.5 &          12.4 &           21.4 &        13.8 &           9.5 &           17.1 &        17.9 &          12.6 &           21.9 &        11.6 &           6.8 &           14.9 &        11.9 &           7.0 &           15.3 \\
		\cmidrule(r){2-4} \cmidrule(r){5-7} \cmidrule(r){8-10} \cmidrule(r){11-13} \cmidrule(r){14-16}
		AVG        & 16.9 & 11.6 & 20.8 	& 13.5 & 9.4 & 16.5 	& 17.5 & 11.8 & 21.7    & 10.6 & 6.1 & 13.7  & 11.0 & 6.5 & 14.1  \\
		\bottomrule
	\end{tabular}
\end{table*}

\begin{table}[h]
	\centering
	\caption{Comparison of our COP RMS error [mm] results and results from studies \cite{s17010075, JUNG201662, Funk2019, AUDU20071115}. N/A - not available, study \cite{Funk2019} does not report values for COP RMS errors for anterior and lateral direction. Results are shown for lateral (Lat.), anterior (Ant.) and total (Tot.) COP RMS error.}
	\label{tab:COPcomp}
	\begin{tabular}{lccc}
		\toprule 
		Study              & Lat. & Ant. & Tot. \\
		\cmidrule(r){2-4}
		Our: LIN; GAM+hist   & 13.0  & 19.6 & 16.7 \\
		Our: LSTM; GAM   & 9.0    & 14.9 & 12.3 \\
		Study \cite{s17010075} & 29 & 45 & 38 \\
		Study \cite{JUNG201662} & 8.4 & 23.1 & 17.4 \\
		Study \cite{Funk2019} & N/A & N/A  & 22-25 \\
		Study \cite{AUDU20071115} & 10 & 45  & 33 \\
		\bottomrule
	\end{tabular}
\end{table}

Performance of our models on both our and the public dataset is given in Tables \ref{tab:intra} and \ref{tab:intra_synt}, which gives the intra-subject RMS Error. We train the models on the first 50\% of each step of the protocol data and test performance on the last 50\%. In general, LSTM performs about 60\% better than the linear model. 

On our dataset, the {lowest RMS error averaged over the subjects (12.3\,mm) is} obtained using the LSTM and complete raw data from gyroscope, accelerometer and magnetometer.  However, removing raw magnetometer data gives marginally worse results by only 7\%. { Best results with the linear model (subject averaged RMS error of 16.7\,mm)} are obtained when using all raw sensor data from IMUs and 0.1\,s of history (10 samples of history). 

Similar conclusions can be drawn for the public dataset. Here, the best result is 10.6\,mm on average, also using the LSTM model. The RMS error of the linear model is 13.5\,mm. The results on the public dataset are broadly comparable to our dataset. Slightly better results (~15\%) are expected, as the underlying GAM data does not come from IMUs, but from a motion tracking system \cite{synthetic}, which typically achieves better accuracy when estimating segment orientation \cite{Caldas2017, armIMUreview}.

Karatsidis \etal \cite{s17010075} used a dynamic model to predict the COP using only IMUs and reported the RMS error of 38\,mm. Ground truth for COP was obtained using force plates. Jung \etal \cite{JUNG201662} developed dynamically adjustable foot-ground contact models for GRF estimation, which also provided COP estimation, and reported RMS error of 17\,mm. Ground truth for COP was obtained using force plate-embedded treadmill system. Funk \etal \cite{Funk2019} presented a method for the COP estimation during Tai Chi movements from video frames using ANN. Ground truth was obtained using insole foot pressure measurement sensors. They reported the RMS error of 22-25\,mm. Audu \etal \cite{AUDU20071115} have developed a three-dimensional biomechanical model of human standing which enabled estimation of COP using optical motion tracking system. Ground truth for COP was obtained using force plates. They reported the RMS error of 33\,mm during various static poses. Studies \cite{s17010075, JUNG201662, AUDU20071115} also report RMS error for anterior COP and lateral COP. In general, both our results and results from literature show that the RMS error is bigger in the anterior direction compared to the lateral direction. Table \ref{tab:COPcomp} summarizes results from literature and our results.

It has been shown that IMU-based systems can in certain circumstances be reliably used instead of camera-based systems for clinical body motion and gait analyses where portability and cost are considered \cite{armIMUreview}. Typically, where high accuracy is needed camera-based systems have the advantage of being able to achieve better accuracy.  The comparison of RMS errors of the studies \cite{s17010075, JUNG201662}, which used solely IMU data, and studies \cite{Funk2019, AUDU20071115}, which used vision-based and optical sensors, shows that IMUs can give comparable results regarding the estimation of the COP. Our methods give comparable and slightly better results compared to studies \cite{s17010075, JUNG201662, Funk2019, AUDU20071115}.

An alternative to the presented method using IMU technology to estimate the COP is the use of foot-pressure sensors for measuring the COP. Foot-pressure sensors are also wearable sensors where subjects can walk freely and are not constrained as with force-plates, optical systems, or treadmills. The main reason for basing our method on the IMU technology is that it can estimate the COP relative to the pelvis, while insoles give COP relative to the foot. {The presented method allows to estimate the gaitogram, which can further be used to extract spatial gait parameters such as step width, step length, single-stance length, single-stance width and stride length
in relation to pelvis position.} Timmermans \etal \cite{Timmermans2019} lists one of the main limitations of the insoles the fact that they can only measure temporal and not spatial gait parameters. Similarly in several studies \cite{JAGOS20103103, Jagos2017,Mohamed2019} pressure sensors in insoles were used to measure temporal parameters, while IMUs (embedded into insoles) were used to estimate spatial parameters. In two studies performed by Jagos \etal \cite{JAGOS20103103, Jagos2017} the IMUs were placed on both feet and studies report difficulties in estimating spatial parameters. Our results show that placement on feet is among the worst positions for estimating the global COP. This shows that insoles are not themselves sufficient for estimating spatial parameters. In some cases a reason to avoid insoles is ergonomical. Some patients use ankle-foot-orthosis resistors, which can complicate the placement of foot-pressure sensors since there might not be enough space due to the orthosis to insert the insoles \cite{AMINIAN2002689}. However, our approach can be extended by fusing both IMU data and foot-pressure sensor data to give a better estimation of COP.
	
Our method using IMU units is not as accurate as an instrumented treadmill. However, it is less expensive and allows subjects to walk freely while providing the same information about COP - a gaitogram. This allows identification of a walking pattern, which is used, for example, in gait rehabilitation in stroke patients. Patients with hemiparesis (as a result of a stroke) almost inevitably experience an asymmetrical gait pattern, which results in a distorted COP ``butterfly'' pattern.

\subsection{Minimizing the time spent in the BART device -- effect of train set size}
We study the dependence of model performance on the size of data used to train the models. We sample a given fraction of all data, train models on it, and predict on the rest of the data\footnote{The train samples are contiguous blocks of data. Taking contiguous data as opposed to completely randomly sampled data avoids over-optimistic results because the train set data points are not too similar/close to the test data points.}. The predictive performance depends on the particularities of the data in the train set. For example, taking 60\,s of standing data for training the models will result in poorer performance than taking 60\,s of walking data. Therefore, the samples are generated multiple times at each train set size. This way the variation in predictive performance given the size of the train set can be estimated.

\begin{figure}[h!]
	\centering
	\includegraphics[width=0.5\columnwidth]{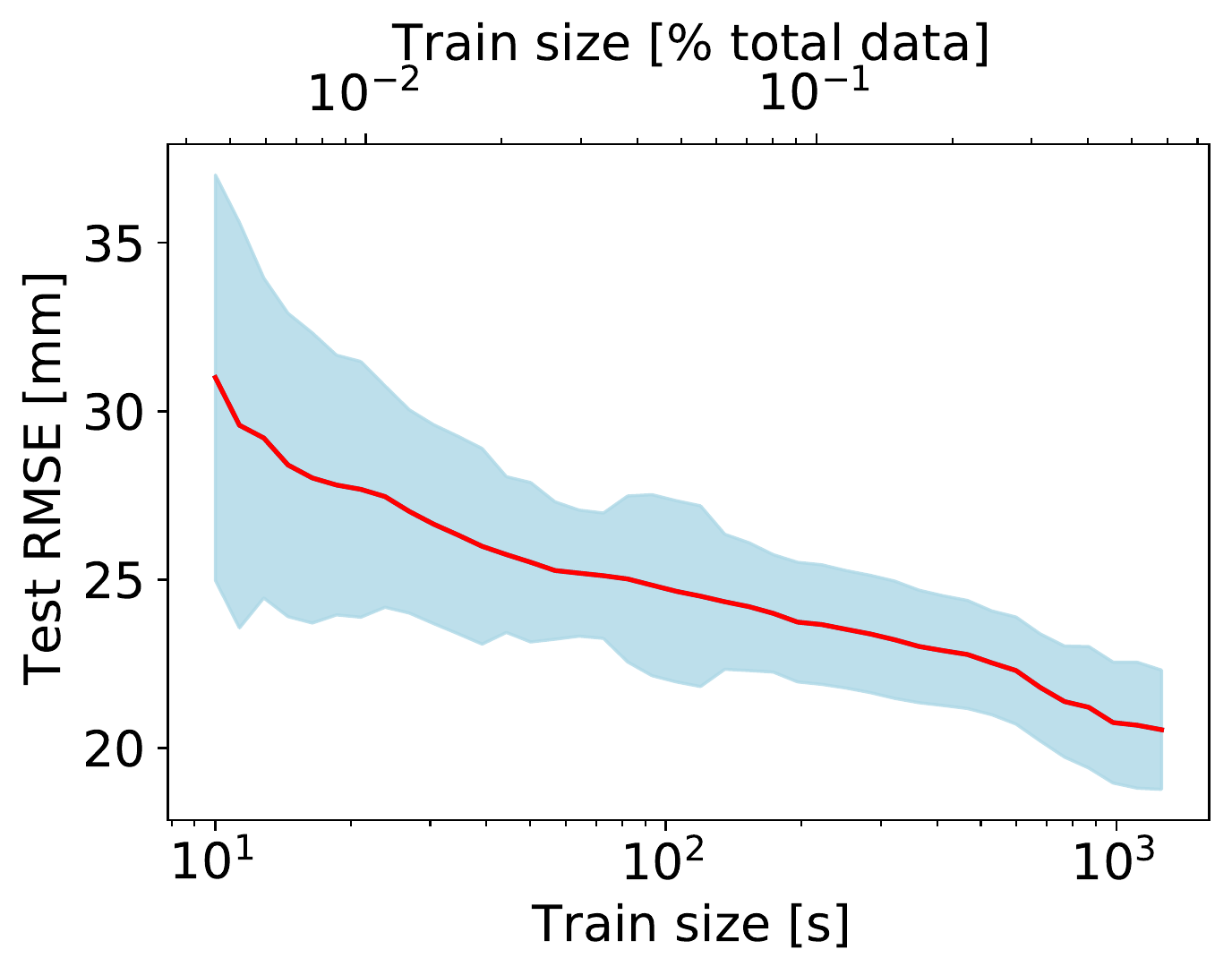}
	\caption{Dependence of the test error on the train set size. Subject S1, linear model. Note the logarithmic scale for the $x$-axis. Shaded regions correspond to one standard deviation about the central line.}
	\label{fig:train_size}
\end{figure}

The results are illustrated in Fig.\,\ref{fig:train_size}. The RMS error for predicting on unseen data decreases exponentially with the size of the train set. For train set sizes above about 100\,s (or about 5\% of the whole measurement protocol) the model performance does not improve significantly. Therefore, each phase of the protocol can be shortened considerably, such that the whole measurement takes no longer than a few minutes, as opposed to about half an hour for the current measurement protocol.

\subsection{Size and robustness of the wearable measurement system}
Walking is highly regular motion and produces regular and periodic signals as measured by each IMU. This means that the IMU signals are strongly correlated when comparing different IMUs or when comparing G, A, M components among themselves. Therefore, removing one or more IMUs from the measurement set-up should not lead to a huge loss of information. Similarly, removing either one or more of GAM components from the input signal should not lead to a large loss of information, and the models should still work well.

Reducing the size of the wearable system by using fewer IMUs contributes to the ease of use and reduces its cost. The magnetic component is most sensitive to environmental disturbances - removing it from the signals measured by IMUs, provided the predictive performance of models does not degrade, contributes to the robustness of the measurement system.

\subsubsection{Effect of reducing the number of IMUs}
\begin{figure}[h!]
	\centering
	\includegraphics[width=0.6\columnwidth]{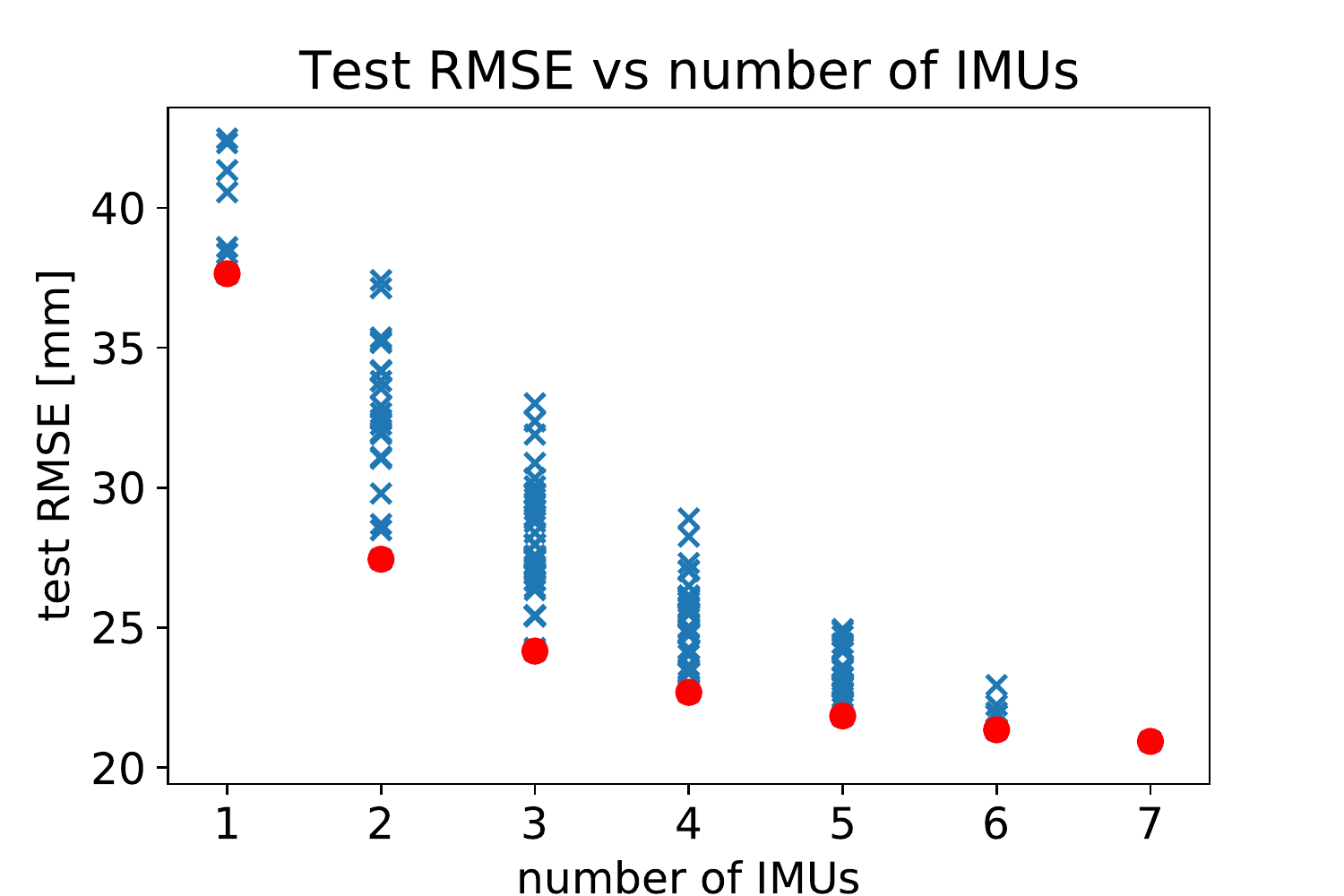}
	\caption{Dependence of the test error on the number of IMUs used. All subjects, linear model. Each point represents a placement configuration of IMUs. Possible locations for IMUs are shown in Figure \ref{fig:IMUsetup}.}
	\label{fig:rmse_n_imu}
\end{figure}

We investigate the dependence of the predictive performance of our models on the number and placement of IMUs used. For a given number of IMUs, we test all possible combinations of placing them to locations labelled 2, 3, 4, 5, 6, 7, and 8 as shown in Fig.\,\ref{fig:IMUsetup}. Since the measurements were made with the full constellation of 7 IMUs, the combination tests are performed by excluding appropriate IMUs from the input data and retraining models on the reduced dataset. There are $2^7 - 1$ possible configuration of IMUs so the exhaustive evaluation of all combinations is not computationally expensive. We record the performance of our models, measured by RMS error, for each configuration of IMUs. The results are presented in Fig.\,\ref{fig:rmse_n_imu} and in Table \ref{tab:imu_placement}. 

This analysis enables us to determine the order of importance of measuring the magneto-inertial data for different body segments. If only one IMU is used, then the placement from the best to the worst is as follows: back, right thigh, left thigh, left shank, right shank, right foot, left foot. This sequence, corresponding to the leftmost set of points plotted in Fig.\,\ref{fig:rmse_n_imu}, is not surprising, as the contributions of individual body segments to the COP are weighted by their mass \cite{s17010075}.

In Fig.\,\ref{fig:rmse_n_imu} we see that there is insignificant gain in model performance above 5 or even only 3 IMUs used in measurements. Table \ref{tab:imu_placement} lists the results for the best and worst locations to place IMUs given the number available, and we can read off that with 3 IMUs the optimal placement is back, and the thighs, whereas with 5 IMUs the optimal placement is on the back, thighs and shanks of the subjects.

A systematic review of gait assessment in Parkinson's disease using IMUs \cite{diseases7010018} reports that the most used setup is with a single sensor, which is most frequently worn on the lower back. Furthermore, the lower back position is the most common position for IMU even where more IMUs are used. Our results show that IMU worn on the lower back is the best placement of IMU if only one IMU is used. Furthermore, IMU on the lower back is present in all best placement configuration, except in configuration where only 2 IMUs are used. In our analysis, the second most important placements are on the thighs which gives the best results when using 2 IMUs. Review \cite{diseases7010018} reports that the next two most common places are on both ankles and both feet. This placement is best for detection of stance and swing events \cite{Caramia2018, Roberts2017, Caldas2017, armIMUreview}, which are one of the most common features that are used in gait analysis. In contrast, our findings show that the placement on feet is the least favorable placement in terms of accuracy for the COP estimation and that IMUs on feet add least to the improvement of accuracy of the COP estimation. Authors of review \cite{diseases7010018} report that they did not find one particular multi-sensor set-up but a wide variety of combinations and that a consensus in this regard is still to be gained in terms of sensor numbers and positions. Authors conclude that placement and multi-sensor set-up would deserve further investigation. Our study and analysis of literature \cite{Caramia2018, Roberts2017, Caldas2017, armIMUreview, diseases7010018} shows that optimal placement depends on particular application and analysis performed.

\begin{table}[h!]
	\centering
	\caption{Best and worst placement configurations of IMUs, given the number of IMUs used. IMU locational labels are as follows (see Figure \ref{fig:IMUsetup}): 2 - back, 3 - right thigh, 4 - right shank, 5 - right foot, 6 - left thigh, 7 - left shank, 8 - left foot}
	\label{tab:imu_placement}
	\begin{tabular}{lll}
		\toprule
		& \multicolumn{2}{c}{IMU placement configuration} \\ \cmidrule(r){2-3}
		No. of IMUs & \multicolumn{1}{c}{Best} & \multicolumn{1}{c}{Worst} \\
		1 &  2 &  8 \\
		2 &  3, 6 &  4, 5 \\
		3 &  2, 3, 6 &  3, 4, 5\\
		4 &  2, 3, 6, 7 &  5, 6, 7, 8\\
		5 &  2, 3, 4, 6, 7 & 2, 5, 6, 7, 8 \\
		6 &  2, 3, 4, 5, 6, 7 &  3, 4, 5, 6, 7, 8\\
		7 &  2, 3, 4, 5, 6, 7, 8 &  \\ 
		\bottomrule
	\end{tabular}
\end{table}

\subsection{Effect of removing magnetometer data from IMU signal.}
Magnetometer data measured by IMUs is the most sensitive to the external changes in the environment. If our model predictions depend on this type of signal, they can be easily corrupted by the proximity of metal and magnets, reducing the robustness of our approach. Therefore, we check the performance of models in normal circumstances (i.e. without magnetic disturbances) when magnetometer data is completely removed from the training set. The results are presented in Table\,\ref{tab:intra}. On average there is an increase in the RMS error without the magnetometer data, however, the relative increase is small (5\%), suggesting that the magnetic component can be safely discarded without the large change in predictive performance.

\subsection{Model transferability}
Constructing models for each subject takes time (the duration of the protocol) and has to be done at the location of the BART device as each subject has to generate enough training data to construct their particular model. After that, the wearable system can be used to deduce the COP position of this subject during walking without the need to use the BART device. 

Measured IMU data is a proxy for the kinematics and dynamics of body segments, which, if known in full detail, determine the COP coordinates. Therefore, despite the specificity of each individual's gait, we expect the models we train on a \emph{set} of subjects to learn the underlying kinematics and dynamics and thus generalise to other subjects, eliminating the need to use the BART device.

To achieve transferability, we need to train the models such that the more general IMU-COP mapping is learned, resulting in models that work well for any/all subjects. The simplest approach is training a \emph{single} model on data from \emph{multiple} subjects. The predictive performance is then evaluated on the data of subjects unseen during training.

The major obstacles to transferability are the different sizes, masses, and moments of inertia of subjects body segments as well as the slightly different placement of IMUs on different subjects. Repeatability of IMU placement is also an issue for measurements on the same subjects but across multiple sessions.

We show that the issue of IMU placement can be alleviated by a short calibration procedure -- subjects stand still for a short time while the IMU and COP coordinate data is measured. This data is then used to retrain the general model to make it more subject-specific. The reason the suggested calibration procedure requires the subjects to stand still is that it can be performed on a force plate, without the need to involve the large BART device. 

The results for our and the public datasets are presented in Tables \ref{tab:transfer} and \ref{tab:transfer_synt}. Two cases are presented: (A) case where no target subject data is used and (B) case where 30\,s of target subject standing data is used to improve the pre-trained models\footnote{For the public dataset we use the data from the static trial}. Again, the LSTM network gives better results. For our dataset, in case (A) the average RMS error is 23.7\,mm, while in case (B) the RMS error is 18.4\,mm. For the public dataset, in case (A) the average RMS error is 21.5\,mm, while in case (B) the RMS error is 17.3\,mm. Additional training of general LSTM network with 30\,s of signals of the target subject improves results of the inter-subject RMS error by about 20\%. 

\begin{table}[h!]
	\centering
	\caption{Inter-subject RMS Error [mm] for the case where no target subject data is used (A) and the case where 60 s of target subject standing data is used to improve pre-trained models (B)}
	\label{tab:transfer}
	\begin{tabular}{lccccccc}
		\toprule
		\multicolumn{1}{c}{} & \multicolumn{2}{c}{Linear model}                                                                       & \multicolumn{2}{c}{LSTM}                                                 \\ \cmidrule(r){2-3} \cmidrule(r){4-5} 
		Subject              & \multicolumn{1}{c}{A} & \multicolumn{1}{c}{B}  & \multicolumn{1}{c}{A} & \multicolumn{1}{c}{B} \\
		S1                   & 29.8                    & 28.0                         & 25.4               & 21.0              \\
		S2                   & 23.6                    & 23.1                         & 21.4                & 20.4               \\
		S3                   & 22.4                    & 22.0                         & 21.8                & 17.5               \\
		S4                   & 26.6                    & 25.4                         & 22.2                & 18.7               \\
		P1                   & 28.9                    & 27.8                         & 23.2                & 15.4               \\
		P2                   & 28.1                    & 24.8                         & 28.1                & 17.5                \\ \cmidrule(r){2-3} \cmidrule(r){4-5} 
		AVG & 26.6 &	25.2 &	23.7 &	18.4  \\
		\bottomrule
	\end{tabular}
\end{table}

\begin{table}[]
	\caption{Inter-subject RMS Error [mm] for the public dataset \cite{synthetic} for the case where no target subject data is used (A) and the case where 60 s of target subject standing data is used to improve pre-trained models (B).}
	\label{tab:transfer_synt}
	\begin{minipage}{0.5\textwidth}
	\begin{tabular}{lccccccc}
		\toprule
		\multicolumn{1}{c}{} & \multicolumn{2}{c}{Linear model}                                                                       & \multicolumn{2}{c}{LSTM}                                                 \\ \cmidrule(r){2-3} \cmidrule(r){4-5} 
		Subject              & \multicolumn{1}{c}{A} & \multicolumn{1}{c}{B}  & \multicolumn{1}{c}{A} & \multicolumn{1}{c}{B} \\
		01      &        16.7 &        16.2 &        15.6 &        13.6 \\
02      &        24.0 &        20.7 &        21.9 &        17.5 \\
03      &        19.1 &        18.1 &        17.1 &        13.5 \\
04      &        18.5 &        18.6 &        19.1 &        17.4 \\
05      &        22.9 &        19.5 &        27.7 &        20.6 \\
06      &        19.2 &        18.0 &        18.9 &        16.7 \\
07      &        21.3 &        19.4 &        16.8 &        15.0 \\
08      &        20.9 &        19.7 &        17.7 &        15.6 \\
09      &        20.4 &        18.6 &        20.0 &        16.8 \\
10      &        20.1 &        18.2 &        18.5 &        15.1 \\
11      &        29.0 &        27.0 &        24.5 &        21.3 \\
12      &        19.9 &        17.9 &        21.8 &        17.6 \\
13      &        24.8 &        20.4 &        23.6 &        18.5 \\
14      &        16.8 &        15.3 &        20.7 &        18.0 \\
15      &        24.9 &        23.2 &        22.1 &        19.0 \\
16      &        19.9 &        18.1 &        17.2 &        14.7 \\
18      &        31.9 &        26.6 &        36.3 &        24.0 \\
\cmidrule(r){2-3} \cmidrule(r){4-5} 
		AVG & 22.3 &	20.2 &	21.5 &	17.3  \\
		\bottomrule
	\end{tabular}
    \end{minipage} \hfill
    \begin{minipage}{0.5\textwidth}	
	\begin{tabular}{lccccccc}
		\toprule
		\multicolumn{1}{c}{} & \multicolumn{2}{c}{Linear model}                                                                       & \multicolumn{2}{c}{LSTM}                                                 \\ \cmidrule(r){2-3} \cmidrule(r){4-5} 
		Subject              & \multicolumn{1}{c}{A} & \multicolumn{1}{c}{B}  & \multicolumn{1}{c}{A} & \multicolumn{1}{c}{B} \\
19      &        19.3 &        17.1 &        16.9 &        12.6 \\
20      &        17.0 &        15.3 &        23.0 &        14.2 \\
21      &        24.0 &        22.5 &        24.3 &        22.0 \\
22      &        19.9 &        18.4 &        18.1 &        16.5 \\
23      &        24.1 &        20.7 &        20.8 &        16.0 \\
24      &        20.5 &        19.8 &        26.0 &        17.4 \\
25      &        19.1 &        17.5 &        17.7 &        15.4 \\
26      &        21.9 &        20.6 &        16.2 &        14.1 \\
27      &        23.2 &        20.7 &        20.9 &        16.8 \\
30      &        17.3 &        15.2 &        12.8 &        11.3 \\
31      &        23.6 &        21.8 &        19.7 &        15.5 \\
33      &        31.4 &        25.2 &        26.7 &        20.7 \\
34      &        28.5 &        27.8 &        29.5 &        24.3 \\
35      &        21.9 &        19.3 &        24.2 &        18.6 \\
38      &        27.5 &        25.3 &        29.0 &        23.4 \\
40      &        27.4 &        23.8 &        25.2 &        18.8 \\
       &          &          &          &        \\ \cmidrule(r){2-3} \cmidrule(r){4-5} 
		AVG & 22.3 &	20.2 &	21.5 &	17.3  \\
		\bottomrule
	\end{tabular}    
    \end{minipage}
\end{table}

\subsection{Study limitations}
The majority of wearable IMU systems including ours might suffer from soft tissue artefacts. Soft tissue artefacts are problematic for all type of sensor systems that require the attachment of part of the sensor system to the subjects, including optical systems, which require the attachment of markers. Karatsidis \etal \cite{s17010075} made a comparison of estimating COP from IMU data and optical motion capture data and has shown that there is only a slight improvement in estimation using optical motion capture system. However, our method makes no assumption about the rigidity of body segments and our models train on raw IMU data. Therefore, the soft tissue effects get to some extent automatically incorporated into our models during learning.

We did not consider all possible statistical models or methods of pre-processing of the input data which might improve the accuracy of the estimated COP. However, this was not the focus of this paper. Similarly, we did not address the issues of dynamic stability or recognition and characterisation of pathological gait, which is our ongoing work stemming from the finding of this contribution.

Our conclusions and statistical `significance' would be stronger if we had a larger number of subjects measured (so far only six). However, we have not observed any large differences among our models of the subjects' gait. We have also evaluated our method on a public dataset\cite{synthetic} restructured to resemble our measurements. We obtained comparable results to our own dataset in terms of RMS error. Therefore, we expect that the addition of more subjects to our study would not change the core of our findings and conclusions.

\section{Conclusions}
\label{sec:conclusion}
We have measured six subjects wearing a constellation of 7 IMUs while walking in an instrumented treadmill. Recorded IMU signals consist of the gyroscope, accelerometer, and magnetometer data whereas the treadmill device records the position of the COP and the centre of the pelvis. 

In this paper we propose a novel approach to modelling the mapping between the raw IMU data and the COP coordinates using only statistical models, without regard to the biomechanical modelling of the human body during walking. We consider a linear model as a baseline as well as a non-linear LSTM neural network to identify the dynamic relationships between the raw sensor data from IMUs and the measured COP. 

We have found that simple statistical models such as linear regression can accurately predict the location of the COP from raw IMU data to better than 1.7\,cm, whereas a simple non-linear LSTM model leads to predictions within 1.2\,cm. We also validated our methods on a public dataset containing gait data of 33 subjects and achieved similar performance. Accuracy of the proposed methods was compared to results from related studies showing that our results are comparable and in some cases better. Accuracy was compared to methods based on kinematic and dynamic models of the human body. Applications of these approaches require a certain level of modelling of biomechanics as well as the data on the body segments of the particular subject (masses, dimensions, and COMs). These are subject-dependent and have to be estimated well for correct modelling. In some cases, kinematic/dynamic approaches also require data from many IMU sensors (for example, 17 sensors were used for data collection in study \cite{s17010075}), which might limit the applicability within a real-life context.

In addition to demonstrating the viability of our approach, we have shown that removing sensor data from magnetometers does not reduce the accuracy substantially, implying that data from gyroscopes and accelerometers is sufficient for estimation of the COP during walking using IMUs. 

We have shown that the measurement protocol in the treadmill device can be as short as a couple of minutes, compared to half an hour used so far, without significant degradation in our model performance.

Furthermore, this study shows that estimating COP is possible using IMUs attached to various segments of the human body. We analysed various configurations of IMUs, in terms of number and placement of sensors. Analysis has shown that the best placement of IMUs for estimating the COP is different from best placement suggested by literature for estimating gait events. This implies that different types of parameters for analyzing gait might require different placements of sensors for optimal estimation.

We have shown that the models trained on one set of subjects generalises and transfers well to a different set of subjects with slightly lower accuracy compared to models specifically trained for a particular subject. However, we expect that with more recorded data and the use of more sophisticated statistical models the accuracy of generic models can be improved, such that the wearable system can be used without the need to calibrate each subject in the instrumented treadmill.

\vspace{6pt} 



\authorcontributions{Conceptualization, J.P. and M.M.; Methodology, J.P, D.K., M.Z., M.M.; Resources, J.P., M.M. and M.Z; Validation and Formal analysis, D.K; Investigation D.K, J.P, M.Z.; Writing—Original draft preparation, D.K and J.P; Writing—review and editing, M.M.; Supervision and Project Administration, M.M.}

\funding{The authors acknowledge the financial support from the Slovenian Research Agency (research core funding No. P2-0228). The authors acknowledge the project (Mechanisms underlying dynamic balancing during human walking, J2-8172) was financially supported by the Slovenian Research Agency. The authors acknowledge the work was supported by the European Community’s H2020 Research and Innovation Programme under grant number 731931 (the CYBERLEGs Plus Plus collaborative project).}

\acknowledgments{The authors thank Jožica Piškur for helping with the measurements.}

\conflictsofinterest{The authors declare no conflict of interest.} 

\abbreviations{The following abbreviations are used in this manuscript:\\

\noindent 
\begin{tabular}{@{}ll}
COP & Centre of pressure\\
COM & Centre of mass\\
GRF & Ground reaction force\\
IMU & Inertial measurement units\\
GAM & Gyroscope, accelerometer, magnetometer \\
BART & Balance assessment robot with an instrumented treadmill\\
ANN & Arteficial neural network\\
LSTM & Long Short-Term Memory neural network\\
RMS & Root mean square \\
DOF & Degrees of freedom \\
\end{tabular}}


\reftitle{References}

\end{document}